\documentclass[twocolumn]{article}

\usepackage{amsthm}
\usepackage{amssymb}
\usepackage{amsmath}
\usepackage{stmaryrd}   
\usepackage{mathtools}

\usepackage{bm}

\usepackage{mathrsfs} 
\DeclareMathAlphabet{\mathpzc}{OT1}{pzc}{m}{it} 
\newcommand{\HilbertSpace}[1]{\mathcal{#1}}

\usepackage{enumitem}
\setlist{
  topsep=1pt,
  itemsep=0pt,
  parsep=1pt
}

\newcommand{\reference}[8]
{
  \noindent
  {#1} {({#6})}: \href{#8}{\it #2}, {#3} {\bf #4} {#5} {#7}
}

\usepackage[safe]{tipa} 
\newcommand{\shape}{
  \raisebox{1pt}{\rm\normalfont\textesh}
}

\def\acts{\raisebox{1.4pt}{\;\rotatebox[origin=c]{90}{$\curvearrowright$}}\hspace{.5pt}}

\usepackage{adjustbox}

\usepackage{cuted}
\usepackage{hyperref}
\hypersetup{
  colorlinks = true,
  allcolors=.  
}

\def\orbisingular{\rotatebox[origin=c]{70}{$\prec$}}
\def\orbisingularG{\raisebox{-3pt}{$\orbisingular^{\hspace{-5.7pt}\raisebox{2pt}{\scalebox{.83}{$G$}}}$}}

\usepackage{tikz}
\usetikzlibrary{cd}

\title{
  Higher Topos Theory in Physics
}

\author{Urs Schreiber${}^\dagger$}

\begin{document}

\setlength{\abovedisplayskip}{2pt}
\setlength{\belowdisplayskip}{2pt}
\setlength{\abovedisplayshortskip}{-10pt}
\setlength{\belowdisplayshortskip}{2pt}

\maketitle

Recall that a {\it category} is a class of objects $X$, $Y$, ... (eg. sets, vector spaces, manifolds, groups, ...) with prescribed sets $\mathrm{Hom}(X,\,Y)$ of (homo)morphisms between them, regarded abstractly as maps $f \,\colon\,X \to Y$ and ultimately defined by their composition law,
\begin{equation}
  \label{CompositionOfMorphisms}
  \begin{array}{c}
  \hspace{-.5cm}
  (\mbox{-})\circ (\mbox{-})
  :
  \!\!
  \begin{tikzcd}
    \mathrm{Hom}(Y,Z)
    \times
    \mathrm{Hom}(X,Y)
    \to
    \mathrm{Hom}(X,Z)
  \end{tikzcd}
  \hspace{-.5cm}
  \\
  \begin{tikzcd}[
    column sep=15pt
  ]
    X
    \ar[
      r,
      shorten=-2pt,
      "{ f }"
    ]
    \ar[
      rr,
      rounded corners,
      to path={
           ([yshift=+00pt]\tikztostart.south)
        -- ([yshift=-10pt]\tikztostart.south)
        --  node[yshift=6pt] {
            \scalebox{.8}{$
              g \circ f
            $}
        }
           ([yshift=-10pt]\tikztotarget.south)
        -- ([yshift=+00pt]\tikztotarget.south)
      }
    ]
    &
    Y
    \ar[
      r,
      shorten=-2pt,
      "{ g }"
    ]
    &
    Z
    \\[-4pt]
  \end{tikzcd}
  \raisebox{4pt}{,}
  \end{array}
\end{equation}
which is required to be associative and unital. The point of category theory is to reason about (functors, natural transformations and then) {\it dualities} in the form of adjunctions.

The most basic understanding of these notions should be sufficient to appreciate this entry.
An introduction to categories aimed at mathematical physicists may be found in [Geroch 1985], for further basic exposition we recommend [Awodey 2006].

\vspace{1cm}

\noindent
\paragraph{Make place for mathematical physics.}
A {\it topos} is a category inside of which the core machinery of mathematics can {\it take place} (whence the name $\tau\mbox{\'o}\pi{o}\varsigma$, for ``place'', plural: $\tau\mbox{\'o}\pi{o}\iota$). For instance, the usual category of sets is a topos, and most of the rigorous mathematics of the 20th century (tacitly) takes place in this default topos of sets.
But there are other toposes. Traditional expositions highlight the {\it petit} toposes of sheaves (see below) on the open subsets of a fixed topological space or on the affine patches of a scheme --- but these traditional examples are {\it not} instructive for the main application of higher topos theory in physics:
Instead, practicing mathematical physicists are mostly working
(mainly unknowingly) inside {\it gros} toposes of sheaves on a category whose objects are {\it all} the {\it probe spaces} on which a given notion of geometry is modeled.
We discuss this by way of the key examples:
\vspace{-1.5cm}
\begin{strip}
\hspace{-7pt}
\def\tabcolsep{5pt}
\begin{tabular}{cc}
  \adjustbox{
    fbox,
    raise=3pt
  }{
  \hspace{-3pt}
  \begin{tikzcd}[
    column sep=-14pt,
    row sep=2pt
  ]
    &[+5pt]
    \scalebox{.7}{
      \def\arraystretch{.9}
      \begin{tabular}{c}
        Cartesian
        \\
        space
      \end{tabular}
    }
    &
    \scalebox{.7}{
      \def\arraystretch{.9}
      \begin{tabular}{c}
        infinitesimal
        \\
        halo
      \end{tabular}
    }
    &
    \scalebox{.7}{
      \def\arraystretch{.9}
      \begin{tabular}{c}
        super
        \\
        space
      \end{tabular}
    }
    &
    \scalebox{.7}{
      \def\arraystretch{.9}
      \begin{tabular}{c}
        higher
        \\
        morphism
      \end{tabular}
    }
    &
    \scalebox{.7}{
      \def\arraystretch{.9}
      \begin{tabular}{c}
        orbi-
        \\
        singularity
      \end{tabular}
    }
    &
    \scalebox{.7}{
      \def\arraystretch{.9}
      \begin{tabular}{c}
        neg-dim
        \\
        sphere
      \end{tabular}
    }
    \\
    \mbox{probe}
    &
    \mathbb{R}^n
    \ar[
      r, phantom, "{ \times }"
    ]
    &
    \mathbb{D}^m_k
    \ar[
      r, phantom, "{ \times }"
    ]
    &
    \mathbb{R}^{0 \vert q}
    \ar[
      r, phantom, "{ \times }"
    ]
    &
    \Delta^{ r }
    \ar[
      r, phantom, "{ \times }"
    ]
    &
    \orbisingularG
    \ar[
      r, phantom, "{ \times }"
    ]
    &
    \mathbb{S}^{-d}
    \\
    \mbox{geometry}
    &
    \scalebox{.7}{
      \def\arraystretch{.9}
      \begin{tabular}{c}
        differential
        \\
        topology
      \end{tabular}
    }
    &
    \scalebox{.7}{
      \def\arraystretch{.9}
      \begin{tabular}{c}
        differential
        \\
        geometry
      \end{tabular}
    }
    &
    \scalebox{.7}{
      \def\arraystretch{.9}
      \begin{tabular}{c}
        super
        \\
        geometry
      \end{tabular}
    }
    &
    \scalebox{.7}{
      \def\arraystretch{.9}
      \begin{tabular}{c}
        homotopy
        \\
        theory
      \end{tabular}
    }
    &
    \scalebox{.7}{
      \def\arraystretch{.9}
      \begin{tabular}{c}
        proper
        \\
        equivariance
      \end{tabular}
    }
    &
    \scalebox{.7}{
      \def\arraystretch{.9}
      \begin{tabular}{c}
        stable
        \\
        homotopy
      \end{tabular}
    }
    \\
    \mathrm{physics}
    &
    \scalebox{.7}{
      \def\arraystretch{.9}
      \begin{tabular}{c}
        fields
      \end{tabular}
    }
    &
    \scalebox{.7}{
      \def\arraystretch{.9}
      \begin{tabular}{c}
        variations
      \end{tabular}
    }
    &
    \scalebox{.7}{
      \def\arraystretch{.9}
      \begin{tabular}{c}
        fermions
      \end{tabular}
    }
    &
    \scalebox{.7}{
      \def\arraystretch{.9}
      \begin{tabular}{c}
        gauge
        \\
        symmetry
      \end{tabular}
    }
    &
    \scalebox{.7}{
      \def\arraystretch{.9}
      \begin{tabular}{c}
        orbi-
        \\
        singularities
      \end{tabular}
    }
    &
    \scalebox{.7}{
      \def\arraystretch{.9}
      \begin{tabular}{c}
        quantum
      \end{tabular}
    }
  \end{tikzcd}
  \hspace{-10pt}
  }
  &
  \begin{minipage}{5.2cm}
    \footnotesize
    The local algebraic coordinate manipulations used so successfully but often informally in physics may be regarded as defining generalized coordinate charts which serve as ``probes'' for the actual global geometric spaces in which physics takes place.  The theory of (higher, {\it gros}) toposes may be understood as making this precise: Probes form a (higher) {\it site} and global spaces form the (higher) {\it sheaf topos} on such a site.
  \end{minipage}
\end{tabular}
\end{strip}

\smallskip

\noindent
\paragraph{Probing space.}
Where a smooth manifold is a set equipped with a smooth structure which is {\it locally diffeomorphic} to Cartesian spaces $\mathbb{R}^n$, we may more generally ask only that a smooth set $X$ be whatever may consistently be {\it probed} by {\it plotting out} Cartesian spaces inside it -- the idea being that such a plot is a smooth map ``$\,\mathbb{R}^n \to X\,$'', only that at this point of bootstrapping $X$ into existence we are yet to say what this means.
But first, for consistency the system of sets of $n$-dimensional plots
\[
  \overset{
    \mathclap{
      \scalebox{.7}{
        \color{gray}
        \def\arraystretch{.9}
        \begin{tabular}{c}
          space
        \end{tabular}
      }
    }
  }{
    {\color{gray} X}
  }
  \;\;:\;\;
  \overset{
    \mathclap{
      \scalebox{.7}{
        \color{gray}
        \def\arraystretch{.9}
        \begin{tabular}{c}
          probe
        \end{tabular}
      }
    }
  }{
    \mathbb{R}^n
  }
  \;\;\mapsto\;\;
  \overset{
    \mathclap{
      \scalebox{.7}{
        \color{gray}
        \def\arraystretch{.9}
        \begin{tabular}{c}
          set of plots
        \end{tabular}
      }
    }
  }{
    \mathrm{Plt}(\mathbb{R}^n, X)
  }
\]
 of $X$ (the latter to be defined thereby) should, clearly, satisfy the following conditions:
\newpage

\noindent
  (1.) {{\bf precomposition of plots} (pre-sheaf condition)}
  \\
  For $\phi \,\in\, \mathrm{Plt}(\mathbb{R}^n, X)$ and smooth $f \,:\, \mathbb{R}^{n'} \to \mathbb{R}^n$, the would-be composition ``\;$\mathbb{R}^{n'} \xrightarrow{f} \mathbb{R}^n \xrightarrow{\phi} X\;$'' should exist as $\phi \!\circ\! f\,\in\, \mathrm{Plt}(\mathbb{R}^{n'}, X)$, such that $\phi \!\circ\! \mathrm{id} = \phi$ and $(\phi \circ f) \circ f' \,=\, \phi \circ (f \circ f')$. In category theoretical language this says that $\mathrm{Plt}(-,X)$ is a {\it presheaf of sets} on the category $\mathrm{CrtSp}$ of smooth Cartesian spaces.

  \noindent
  (2.) {\bf gluing of plots} (sheaf condition)
  Given an open cover $\big\{ U_i \xhookrightarrow{ \iota_i } \mathbb{R}^n \big\}_{i \in I}$ which is {\it differentiably good} --- meaning that the patches $U_i$ and their intersections $U_i \cap U_j$ are all diffeomorphically identified with $\mathbb{R}^n$ --- those $I$-tuples of plots $\phi_i$ by the $U_i$ which coincide on the overlaps $U_i \cap U_j$ should be in natural bijection with the global plots by the full $\mathbb{R}^n$:
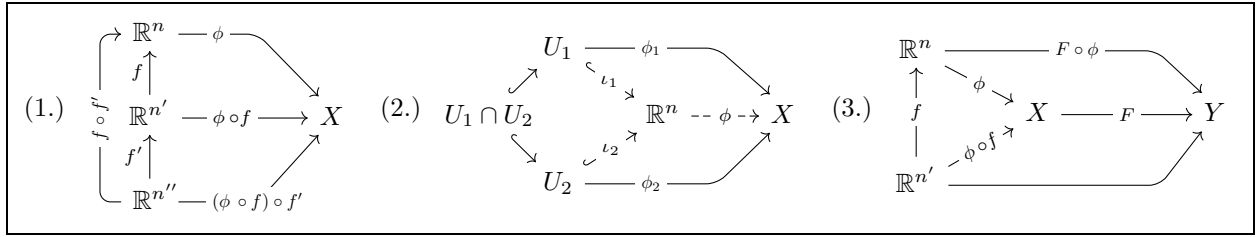
  \[
    \begin{tikzcd}[
      column sep=6pt,
      row sep=-4pt
    ]
      \mathrm{Plt}(\mathbb{R}^n, X)
      \ar[
        r,
        shorten=-2pt,
        "{ \sim }"{sloped}
      ]
      &
      \Big\{
        \big(
          \phi_i
          \in
          \mathrm{Plt}(U_i, X)
        \big)_i
        \,\big\vert\,
        \;\,
        \underset{
          \mathclap{
          \scalebox{.5}{$U_i \cap U_j$}
          }
        }{\forall}
        \;
        \phi_i = \phi_j
      \Big\}
      \\
      \phi
      \ar[
        r,
        |->,
        shorten=20pt
      ]
      &
      \big(
        \phi \circ \iota_i
      \big)_{i \in I}
    \end{tikzcd}
  \]
  In topos-theoretic language, this says that the presheaf $\mathrm{Plt}(-,X)$ must be a {\it sheaf} with respect to the {\it coverage} (aka: {\it Grothendieck pre-topology}) of differentiably good open covers.

\noindent
(3.) {\bf postcomposition of plots} (sheaf morphism)
A {\it smooth map} $F \,:\, X \to Y$ between such smooth sets should be whatever consistently takes plots $\phi \,\in\, \mathrm{Plt}(\mathbb{R}^n, X)$ to their would-be composites\\ ``$\;\mathbb{R}^n \xrightarrow{\phi} X \xrightarrow{ F } Y\;$'', being  $F \circ \phi \,\in\, \mathrm{Plt}(\mathbb{R}^n, Y)$, \\ consistency requiring that for $f \,:\, \mathbb{R}^{n'} \to \mathbb{R}^n$ smooth we have $(F \circ \phi) \circ f \,=\, F \circ (\phi \circ f)$.
In category/topos-theoretic laguage this says that smooth maps $F$ are the {\it morphisms} in the topos of sheaves (of plots) on the category $\mathrm{CrtSp}$ regarded as a ``site'' (a category equipped with a notion of gluing of its objects) via the above coverage.

\begin{strip}
  \adjustbox{
    fbox
  }{
  (1.)
  \begin{tikzcd}[
    row sep=15pt
  ]
    \mathbb{R}^n
    \ar[
      drr,
      "{  }",
      rounded corners,
      to path={
           ([xshift=0pt]\tikztostart.east)
        -- node {
             \scalebox{.7}{\colorbox{white}{$
               \phi
             $}}
           }
           ([xshift=30pt]\tikztostart.east)
        -- ([xshift=-05pt]\tikztotarget.north)
      }
    ]
    \\
    \mathbb{R}^{n'}
    \ar[
      u,
      shorten <=-2pt,
      "{ f }"{pos=.45}
    ]
    \ar[
      rr,
      "{
        \phi\, \circ f
      }"{description, pos=.4}
    ]
    &&
    X
    \\
    \mathbb{R}^{n\mathrlap{{}''}}
    \ar[
      u,
      shorten <=-2pt,
      "{ f' }"{pos=.45}
    ]
    \ar[
      urr,
      "{  }",
      rounded corners,
      to path={
           ([xshift=0pt]\tikztostart.east)
        -- node[xshift=15pt] {
             \scalebox{.7}{\colorbox{white}{$
        (\phi \, \circ f) \circ f'
             $}}
           }
           ([xshift=30pt]\tikztostart.east)
        -- ([xshift=-05pt]\tikztotarget.south)
      }
    ]
    \ar[
      uu,
      rounded corners,
      to path={
           ([xshift=-00pt]\tikztostart.west)
        -- ([xshift=-8pt]\tikztostart.west)
        -- node[sloped] {
          \scalebox{.7}{
            \colorbox{white}{$
              f \circ f'
            $}
          }
        }
           ([xshift=-8pt]\tikztotarget.west)
        -- ([xshift=-00pt]\tikztotarget.west)
      }
    ]
  \end{tikzcd}
  \hspace{.1cm}
  (2.)
  \begin{tikzcd}[
    column sep=15pt,
    row sep=10pt
  ]
    &[-20]
    U_1
    \ar[
      dr,
      hook,
      "{ \iota_1 }"{description}
    ]
    \ar[
      drrr,
      rounded corners,
      to path={
           ([xshift=0pt]\tikztostart.east)
        -- node {
           \scalebox{.7}{
             \colorbox{white}{$
               \phi_1
             $}
           }
        }
           ([xshift=50pt]\tikztostart.east)
        -- ([xshift=-05pt]\tikztotarget.north)
      }
    ]
    &[+4pt]
    &[-4pt]
    \\
    U_1 \cap U_2
    \ar[ur, hook]
    \ar[dr, hook]
    &&
    \mathbb{R}^n
    \ar[
      rr,
      dashed,
      "{ \phi }"{description}
    ]
    &&
    X
    \\
    &
    U_2
    \ar[
      ur,
      hook,
      "{ \iota_2 }"{description}
    ]
    \ar[
      urrr,
      rounded corners,
      to path={
           ([xshift=0pt]\tikztostart.east)
        -- node {
           \scalebox{.7}{
             \colorbox{white}{$
               \phi_2
             $}
           }
        }
           ([xshift=50pt]\tikztostart.east)
        -- ([xshift=-05pt]\tikztotarget.south)
      }
    ]
  \end{tikzcd}
  \hspace{.1cm}
  (3.)
  \begin{tikzcd}[
    column sep=20pt,
    row sep=10pt
  ]
    \mathbb{R}^n
    \ar[
      dr,
      "{\phi}"{description}
    ]
    \ar[
      drrrr,
      "{  }",
      rounded corners,
      to path={
           ([xshift=0pt]\tikztostart.east)
        -- node[xshift=10pt] {
             \scalebox{.7}{\colorbox{white}{$
               F \circ \phi
             $}}
           }
           ([xshift=80pt]\tikztostart.east)
        -- ([xshift=-05pt]\tikztotarget.north)
      }
    ]
    &[+5pt]
    &[-5pt]
    &[-5pt]
    \\
    &
    X
    \ar[
      rrr,
      "{ F }"{description}
    ]
    &&&
    Y
    \\
    \mathbb{R}^{n'}
    \ar[
      ur,
      "{
        \phi \, \circ f
      }"{description, sloped}
    ]
    \ar[
      uu,
      "{ f }"{description}
    ]
    \ar[
      urrrr,
      "{  }",
      rounded corners,
      to path={
           ([xshift=0pt]\tikztostart.east)
        -- ([xshift=80pt]\tikztostart.east)
        -- ([xshift=-05pt]\tikztotarget.south)
      }
    ]
  \end{tikzcd}
  }
\end{strip}

This is all fairly self-evident, and yet  it means that smooth sets form the {\it sheaf topos} on the {\it site} of smooth Cartesian spaces
over the {\it base topos} of sets:
\begin{equation}
  \label{ToposOfSmthSets}
  \def\arraystrech{1.8}
  \begin{array}{l}
  \mathrm{SmthSet}
  \\
  :=
  \underset{
    \mathclap{
      \raisebox{-1pt}{
        \scalebox{.7}{
          \color{gray}
          with gluing condition
        }
      }
    }
  }{
  \mathrm{Sh}(
    \mathrm{CrtSp}
    ,
    \mathrm{Set}
  )
  }
  \hookrightarrow
  \underset{
    \mathclap{
      \raisebox{-1pt}{
        \scalebox{.7}{
          \color{gray}
          without gluing condition
        }
      }
    }
  }{
  \mathrm{PSh}(
    \mathrm{CrtSp}
    ,
    \mathrm{Set}
  )
  \,.
  }
  \end{array}
\end{equation}

For example, a Cartesian space $\mathbb{R}^n$ (and generally a smooth manifold $M$) becomes a smooth set by taking its plots to be the ordinary smooth functions into it:
\begin{equation}
  \label{ExternalYonedaEmbeddingIntoSmthSets}
  \hspace{-2cm}
  \adjustbox{
    raise=-6pt
  }{
  \begin{tikzcd}[
    column sep=2pt,
    row sep=0pt
  ]
    \mathrm{CrtSp}
    \ar[
      r,
      hook
    ]
    &[10pt]
    \mathrm{SmoothMfd}
    \ar[
      r,
      hook
    ]
    &[-10pt]
    \mathrm{SmthSet}
    \\
    &
    M
    \ar[
      r, |->
    ]
    &
    \mathrm{Plt}(-,M)
    :=
    C^\infty(-,M)
    \,.
  \end{tikzcd}
  }
  \hspace{-2cm}
\end{equation}
This means in particular that now we {\it do} have a notion of smooth maps $\mathbb{R}^n \to X$ from a Cartesian space into any smooth set. Consistency of the above ``bootstrap''-definition now demands that the prescribed plots naturally coincide with these actual smooth maps. This crucial demand happens to be satisfied by the general fact which in category theory is famous as the {\bf Yoneda lemma} (the full inclusion \eqref{ExternalYonedaEmbeddingIntoSmthSets} being the {\it Yoneda embedding} that it implies):
\begin{equation}
  \label{YonedaLemma}
  \begin{tikzcd}[
    column sep=6pt,
    row sep=0pt
  ]
    \mathrm{Hom}_{\mathrm{SmthSet}}
    \big(
      \mathbb{R}^n
      ,\,
      X
    \big)
    \ar[
      rr,
      "{\sim}"
    ]
    &&
    \mathrm{Plt}(\mathbb{R}^n,X)
    \\
    F &\mapsto&
    F \circ \mathrm{id}_{\mathbb{R}^n}
  \end{tikzcd}
\end{equation}

Another basic result of topos theory gives that a smooth set is determined already by the {\it germs} of its plots, i.e. their equivalence classes under restriction to smaller neighbourhoods of any point (say $0 \in \mathbb{R}^n$):
\begin{equation}
  \label{GermsOfPlots}
  \hspace{-5pt}
  \adjustbox{raise=-15pt}{$
  \def\arraystretch{1.2}
  \begin{array}{l}
  \mathrm{PltGrm}(\mathbb{R}^{n}, X)
  \\
  :=
  \mathrm{Plt}(\mathbb{R}^{n}, X)
  \big/
  \big(
    \phi \sim \phi'
    \mbox{\;iff\;}
    \;\;
    \underset{
      \mathclap{
      \;\,
      \mathbb{R}^n
      \underset{
        \mathclap{
        \mathrm{opn}
        \atop
        { 0 \in \iota(\mathbb{R}^n) }
        }
      }{
      \overset{\iota}{\hookrightarrow}
      }
      \mathbb{R}^n
      )
      }
    }{\exists}
    \;\;
    (
      \phi \circ \iota
      \,=\,
      \phi' \circ \iota
  \big)
  \,.
  \hspace{-10pt}
  \end{array}
  $}
\end{equation}
\vspace{-5pt}

\noindent
Namely, maps $F : X \to Y$ in $\mathrm{PSh}(\mathrm{CrtSp}, \mathrm{Set})$
restrict to germs of plots
\begin{equation}
  \label{MapsOnGermsOfPlots}
  F \circ
  :
  \mathrm{PltGrm}(
    \mathbb{R}^n
    ,\,
    X
  )
  \to
  \mathrm{PltGrm}(
    \mathbb{R}^n
    ,\,
    X'
  )
\end{equation}
and the {\it localization} $L$ at (forcing the invertibility of) the {\it local isomorphisms}
\[
  \begin{tikzcd}[
    column sep=6pt
  ]
    X
    \ar[
      r,
      shorten=-3pt,
      "{ \mathrm{liso} }"{yshift=1pt},
      "{
        \scalebox{.7}{
          \color{gray}
          \def\arraystretch{.85}
          \begin{tabular}{c}
            local
            \\
            isomorphism
          \end{tabular}
        }
      }"{swap, yshift=-2pt}
    ]
    &
    Y
  \end{tikzcd}
  \hspace{-12pt}
  \Leftrightarrow
  \;\;
  \underset{
    n
  }{\forall}
  \begin{tikzcd}[
    column sep=5pt
  ]
    \mathrm{PltGrm}\big(
      \mathbb{R}^n
      ,
      X
    \big)
    \ar[
      r,
      shorten=-3pt,
      "{
        \mathrm{iso}
      }"{yshift=1pt},
      "{
        \scalebox{.7}{
          \color{gray}
          \def\arraystretch{.85}
          \begin{tabular}{c}
            isomorphism
            \\
            (bijection)
          \end{tabular}
        }
      }"{swap, yshift=-4pt}
    ]
    &
    \mathrm{PltGrm}\big(
      \mathbb{R}^n
      ,
      Y
    \big)
  \end{tikzcd}
\]
\vspace{-.35cm}

\noindent
yields an equivalent category:
\begin{equation}
  \label{SmthSetsAsLocalizationOfPresheaves}
  \mathrm{SmthSet}
  \;\simeq\;
  L^{\!\mathrm{liso}}
  \mathrm{PSh}\big(
    \mathrm{CrtSp}
    ,
    \mathrm{Set}
  \big)
  \,.
\end{equation}

This reflects the intuition that arbitrarily small probes (of any dimension) are sufficient for exploring the smooth structure of a space.

\smallskip

\noindent
\paragraph{Toposes as categories of probeable spaces.}
In conclusion so far, it is remarkable how the fundamental concepts of topos theory naturally align with physical intuition of ``space'' as whatever is witnessed by ``probing'' it (cf. the notion of {\it probe branes} in string theory, whose worldvolumes play much the role of the above plots, now for probing ``D-geometry'').

In fact, that is the definition:
A {\it topos} (here short for: {\it Grothendieck topos}, as usual) is a category $\mathcal{T}$ for which:

\noindent
(1.) there exists a category $S$ consisting of a set (instead of a proper class) of ``probe''-objects, which is a {\it site} in that it is

\noindent
(2.) equipped with a consistent notion ({\it coverage} or {\it Grothendieck pre-topology}) of what it means for any such probe object to be {\it covered} by other probes,

\noindent
(3.)
such that $\mathcal{T}$ is equivalently the category of objects consistently probeable by these probe objects,
in immediate generalization of the example \eqref{ToposOfSmthSets}, called the category of sheaves on $S$:
\begin{equation}
  \label{SheafTopos}
  \hspace{-11pt}
  \scalebox{.8}{
    \color{gray}
    \def\arraystretch{.85}
    \begin{tabular}{c}
      (Grothendieck-)
      \\
      topos
    \end{tabular}
  }
  \hspace{-3pt}
  \mathcal{T}
  \;\simeq\;
  \mathrm{Sh}(S, \mathrm{Set})
  \hspace{-6pt}
  \scalebox{.8}{
    \color{gray}
    \def\arraystretch{.9}
    \begin{tabular}{c}
      category of
      \\
      sheaves on site $S$.
    \end{tabular}
  }
\end{equation}

\smallskip

\noindent
\paragraph{Where field spaces take place.}
As a first example of the power of topos theory in physics, notice that much of the subtlety of field theory in contrast to point mechanics goes back to the fact that physical field configurations $\Phi$ are (smooth) {\it maps} from a spacetime manifold $X$ to some coefficient space
\begin{equation}
  \label{AFieldAsAMap}
  \Phi
  \,:\,
  X \to F
  \,,
\end{equation}
so that spaces of field histories are {\it mapping spaces}  (generally: spaces of sections of $F$-fiber bundles over $X$). These are at best infinite-dimensional Fr{\'e}chet manifolds (under the unrealistic assumption that $X$ is compact) and in general fall entirely outside the scope of traditional differential geometry.

In contrast,
the topos of smooth sets -- like every topos -- is {\it cartesian closed}, meaning that

\noindent
(1.) Cartesian products exist, immediately so by \eqref{YonedaLemma}:
\[
  \mathrm{Plt}\big(
    \mathbb{R}^n
    ,
    X \times Y
  \big)
  \,:=\,
  \mathrm{Plt}\big(
    \mathbb{R}^n
    ,
    X
  \big)
  \times
  \mathrm{Plt}\big(
    \mathbb{R}^n
    ,
    Y
  \big)
  \,,
\]

\noindent
(2.)
there is guaranteed to be a smooth set
\[
  \mathrm{Fields}
    :=
  \mathrm{Maps}(X,F)
\]
of smooth maps (between any two smooth sets $X$, $F$) such that there are natural isomorphisms
$$
  U \to \mathrm{Map}(X,F)
  \hspace{.4cm}
  \leftrightarrow
  \hspace{.4cm}
  U \times X
  \to
  F
  \,.
$$
Concretely, plotting out a probe $U$ inside the space of maps from $X \to F$ --- which one may think of as $X$-parameterized elements of $F$ --- should just be a $U$-parameterized family of such maps, hence a $U \times X$-parameterized family of elements of $F$:
\begin{equation}
  \label{PlotsOfMappingSpace}
  \def\arraystrech{1.3}
  \def\arraycolsep{4pt}
  \begin{array}{rlc}
  \mathrm{Plt}\big(
    U
    ,\,
    \mathrm{Maps}(X,F)
  \big)
  &
  :=
  &
  \mathrm{Hom}\big(
     U \times X
     ,\,
     F
  \big)
  \\
  \begin{tikzcd}[
    column sep=13pt,
    row sep=-2pt
  ]
    U
    \ar[r]
    &
    \mathrm{Fields}
    \\
    u
    \ar[r, phantom, "{ \mapsto }"]
      &
    \Phi_u
  \end{tikzcd}
  &\leftrightarrow&
  \begin{tikzcd}[
    column sep=4pt,
    row sep=-2pt
  ]
    U \times X
    \ar[
      r
    ]
    &
    F
    \\
    (u,x)
    \ar[
      r,
      phantom,
      "{ \mapsto }"
    ]
    &
    \Phi_u(x)
  \end{tikzcd}
  \end{array}
\end{equation}
This most simple and intuitively transparent  prescription {\it defines} the mapping space $\mathrm{Maps}(X,F)$ as a smooth set,
and yet subsumes all the traditional definitions whenever they happen to be applicable:

Namely, also infinite-dimensional Fr{\'e}chet manifolds are faithfully included among smooth sets ---  via the analogous formula \eqref{ExternalYonedaEmbeddingIntoSmthSets}, as are
{\it diffeological spaces} \eqref{PlotsOfConcreteSmthSet}, cf. \hyperlink{FigureS}{\it Figure 1.} --- and when the mapping space exists in these subcategories it agrees with \eqref{PlotsOfMappingSpace}.

\smallskip
\noindent
\paragraph{Where anomaly polynomials take place.}
Contrary to tradition in differential geometry, these smooth sets are {\it not defined} as underlying sets equipped with smooth structure, though this case is subsumed: A smooth set $X$ is {\it concrete} (as a sheaf on $\mathrm{CrtSp}$) -- also called a {\it diffeological space} -- if there exists a plain set $X_{s}$ such that the plots of $X$ are natural subsets of the maps of plain sets into $X_s$:
\begin{equation}
  \label{PlotsOfConcreteSmthSet}
  \begin{tikzcd}
    \mathrm{Plt}(\mathbb{R}^n, X)
    \ar[r, hook]
    &
    \mathrm{Hom}_{\mathrm{Set}}\big(
      \mathbb{R}^n
      ,\,
      X_s
    \big)
    \,.
  \end{tikzcd}
\end{equation}
For example, smooth manifolds \eqref{ExternalYonedaEmbeddingIntoSmthSets} are among concrete smooth sets, but also (Delta-generated) topological spaces $\mathrm{X} \in \mathrm{DTopSp}$ are {\it faithfully} subsumed, via
\begin{equation}
  \begin{tikzcd}[
    row sep=-3pt,
    column sep=10pt
  ]
    \mathrm{DTopSp}
    \ar[
      r,
      hook
    ]
    &
    \mathrm{DifflgSp}
    \ar[
      r,
      hook
    ]
    &
    \mathrm{SmthSet}
    \\
    \mathrm{X}
    \ar[
      rr,
      phantom,
      "{ \mapsto }"{pos=.2}
    ]
    &&
    \mathclap{
    \hspace{-13pt}
    \mathrm{Plt}(-,\mathrm{X})
    :=
    C^0(-,\mathrm{X})
    \mathrlap{\,.}
    }
  \end{tikzcd}
\end{equation}

Among {\it non}-concrete smooth sets are the ``smooth classifying spaces'' $\boldsymbol{\Omega}_{\mathrm{dR}}^p$, of differential $p$-forms:
\[
  \hspace{-6pt}
  \mathrm{Plt}\big(
    \mathbb{R}^n
    ,\,
    \boldsymbol{\Omega}^p_{\mathrm{dR}}
  \big)
  \;\;:=\;\;
  \Omega^p_{\mathrm{dR}}(\mathbb{R}^n)
  \scalebox{.65}{
    \color{gray}
    \def\arraystretch{.85}
    \begin{tabular}{c}
      set of smooth
      \\
      differential $p$-forms
      \\
      on Cartesian space.
    \end{tabular}
  }
  \hspace{-6pt}
\]
These are classifying in that smooth maps from a smooth manifold $X$ into them are in natural bijection with smooth differential forms on $X$:
\[
  \mathrm{Hom}\big(
    X
    ,\,
    \boldsymbol{\Omega}^p_{\mathrm{dR}}
  \big)
  \;\;\simeq\;\;
  \Omega^p_{\mathrm{dR}}(X)
  \,.
\]
Remarkably, if $X$ is an $n$-manifold, then the smooth mapping space \eqref{PlotsOfMappingSpace} into, say, $\boldsymbol{\Omega}^{n+2}_{\mathrm{dR}}$ is still non-trivial: It contains differential forms that appear only {\it in families} parameterized by some manifold $U$:
$$
  U \to \mathrm{Maps}(X,\boldsymbol{\Omega}^p_{\mathrm{dR}})
  \;\;
  \leftrightarrow
  \;\;
  \omega_U \,\in\,
  \Omega^{n+2}_{\mathrm{dR}}(X \times U)
  \,.
$$
This is the case for Green-Schwarz-type anomaly polynomials $I_{n+2}  \,=\, j_{n-k}^{\mathrm{el}} \wedge j_{k+2}^{\mathrm{mag}}$, whose mathematical home, traditionally left vague, is the following diagram of smooth sets (the integral assuming compact support, as usual):
\[
  \begin{tikzcd}[
    column sep=3pt,
    row sep=0pt
  ]
    \mathrm{Map}(X^n,F)
    \ar[
      rr,
      "{
        I_{{}_{n+2}}
      }"
    ]
    \ar[
      rrrr,
      rounded corners,
      to path={
           ([xshift=-10pt,yshift=+00pt]\tikztostart.north)
        -- ([xshift=-10pt,yshift=+9pt]\tikztostart.north)
        -- node {
          \scalebox{.7}{
            \colorbox{white}{
              \color{gray}
              curvature of anomaly line bundle on field space
            }
          }
        }
           ([yshift=+08pt]\tikztotarget.north)
        -- ([yshift=+00pt]\tikztotarget.north)
      }
    ]
    &&
    \mathrm{Map}(X^n,\boldsymbol{\Omega    }^{n+2}_{\mathrm{dR}})
    \ar[
      rr,
      "{ \int_{X^n} }"
    ]
    &&
    \boldsymbol{\Omega}^2_{\mathrm{dR}}
    \\[-7pt]
    (
      \overset{
        \mathclap{
          \raisebox{1pt}{
            \scalebox{.7}{
              \color{gray}
              probe
            }
          }
        }
      }{
        U
      }
      \times
      X^n
        \xrightarrow{\Phi} F
    )
    \ar[
      rr,
      phantom,
      "{\mapsto}"
    ]
    &&
    I_{n+2}(\Phi)
    \ar[
      rr,
      phantom,
      "{\mapsto}"
    ]
    &&
    \int_{X^n}
    I_{{}_{n+2}}(\Phi)
    \\[-6pt]
    \mathclap{
    \scalebox{.7}{
      \color{gray}
      \def\arraystretch{.9}
      \begin{tabular}{c}
        $U$-parameterized
        \\
        family of fields
      \end{tabular}
    }
    }
    &&
    \mathclap{
    \scalebox{.7}{
      \color{gray}
      \def\arraystretch{.9}
      \begin{tabular}{c}
        anomaly polynomial
        \\
        in
        $\Omega^{n+2}(X^n \times U)$
      \end{tabular}
    }
    }
    &&
    \mathclap{
    \scalebox{.7}{
      \color{gray}
      \def\arraystretch{.9}
      \begin{tabular}{c}
        local anomaly
        \\
        in
        $\Omega^{2}(U)$
      \end{tabular}
    }
    }
  \end{tikzcd}
\]
\vspace{-.5cm}

\noindent
\paragraph{Constructivism and instanton sectors.}
A fundamental result of (``elementary'') topos theory is that mathematical definitions and theorems may be transported from plain sets to  any other topos -- if only they are {\it constructive}, meaning essentially that they do not invoke the usual ``axiom of choice''. This is not mysterious but another example of physical intuition aligning with topos theory:

Namely, in the default topos of sets the axiom of choice says equivalently that every surjective map $\begin{tikzcd}[sep=10pt]E \ar[r, ->>] & B\end{tikzcd}$ (``epimorphism'') admits a {\it choice} $b \mapsto \sigma(b) \,\in\, E_b$ of elements in the fiber $E_b$ over each point $b \in B$ in the base, forming a commuting diagram of this form:
\[
\begin{tikzcd}[
  sep=15pt,
  row sep=8pt
]
  && E
     \ar[
       dd,
       ->>,
     ]
  \\
  \\
  B
    \ar[rr, equals]
    \ar[
      uurr,
      dashed,
      "{ \exists\,  ? \, \sigma }",
      "{
        \scalebox{.7}{
          \color{gray}
          a choice
        }
      }"{swap, sloped}
   ]
  &&
  B
\end{tikzcd}
\]
Whatever one may think of this axiom in the case of plain sets, it is clearly {\it un}justified for smooth sets, where a surjection as above is a {\it smooth bundle}, such as a fiber bundle or a principal bundle. Even if we assume (as one usually does) that we can choose elements $\sigma(b)$ in each fiber of the bundle separately, for a non-trivial principal bundle there is in general no way to make theses choices {\it smoothly} (or even continuously) to arrange into a smooth map $\sigma \,\colon\, B \to E$ as above:

Therefore the failure of the axiom of choice in the topos of smooth sets is, in a sense, the {\it reason} why in physics one sees crucial phenomena like flux quantization, soliton/instanton sectors or fermionic anomalies -- all of which reflect the existence of non-trivial fiber bundles.

 Hence reasoning in mathematical physics is naturally reflected in constructive mathematics,
and toposes are essentially the possible {\it models} of such constructive or {\it physical} reasoning.
There are many topos models relevant for the discussion of physics:

\begin{strip}
  \hypertarget{FigureS}{}
  \hspace{10pt}
  \def\arraystretch{1}
  \begin{tabular}{l}
  \adjustbox{
    scale=.95,
    fbox
  }{
  \begin{tikzcd}[
    column sep=8pt,
    row sep=10pt
  ]
    &&
    \mathrm{DTopSp}
    \ar[
      drr,
      hook,
      end anchor={[yshift=2pt]}
    ]
    \\[-28pt]
    \mbox{SmthMfd}
    \ar[rr, hook]
    \ar[d, hook]
    &&
    \mbox{FrchtMfd}
    \ar[rr, hook]
    &&
    \mbox{DifflgSp}
    \ar[rr, hook]
    &&
    \overset{
      \mathclap{
        \hspace{-80pt}
        \raisebox{6pt}{
          \scalebox{.7}{
            \color{gray}
            \def\arraystretch{.9}
            \begin{tabular}{c}
              ungauged bosonic
              \\
              field spaces
              take place here
            \end{tabular}
          }
        }
      }
    }{
      \mbox{SmthSets}
    }
    \ar[rr, hook]
    \ar[
      ddd,
      dashed,
      gray,
      shorten <=-11pt,
      shorten >=-20pt,
      -,
      shift right=44pt,
      "{
        \scalebox{.7}{
          \color{gray}
          \begin{tabular}{l}
            (higher)
            \\
            toposes $\longrightarrow$
          \end{tabular}
        }
      }"{pos=1.12, xshift=-6pt}
    ]
    &&
    \overset{
      \mathclap{
        \raisebox{6pt}{
          \scalebox{.7}{
            \color{gray}
            \def\arraystretch{.9}
            \begin{tabular}{c}
              ungauged fermionic
              \\
              field spaces
              take place here
            \end{tabular}
          }
        }
      }
    }{
      \mbox{SupSmthSets}
    }
    &[-40pt]&[-10pt]
    \\[-13pt]
    \mbox{LieGrpd}
    \ar[rrrr, hook]
    \ar[
      drr,
      shorten <=-2pt,
      hook
    ]
    &&
    &&
    \mbox{DifflgGrpd}
    \ar[rr, hook]
    &&
    \overset{
      \mathclap{
        \raisebox{6pt}{
          \scalebox{.7}{
            \color{gray}
            \def\arraystretch{.9}
            \begin{tabular}{c}
              field space of pure
              \\
              Yang-Mills takes place here
            \end{tabular}
          }
        }
      }
    }{
      \mbox{SmthGrpd}
    }
    \ar[
      drr,
      shorten >= -0pt,
      hook
    ]
    \\[-15pt]
    {}
    &&
    \underset{
      \mathclap{
        \raisebox{-6pt}{
        \scalebox{.7}{
          \color{gray}
          \def\arraystretch{.9}
          \begin{tabular}{c}
            BRST complex of
            \\
            QCD takes place here
          \end{tabular}
        }
        }
      }
    }{
      \mathrm{SupLieGrpd}
    }
    \ar[rrrrrr, hook]
    &&&&&&
    \overset{
      \mathclap{
        \raisebox{6pt}{
          \scalebox{.7}{
            \color{gray}
            \begin{tabular}{c}
              field space of
              \\
              QCD takes place here
            \end{tabular}
          }
        }
      }
    }{
      \mathrm{SupSmthGrpd}
    }
    \ar[dr, hook]
    \\[-20pt]
    &&&&&
    {}
    &
    {}
    &
    {}
    &
    {}
    &
    \underset{
      \mathclap{
        \raisebox{-6pt}{
          \scalebox{.7}{
            \color{gray}
            \begin{tabular}{c}
            field space of higher
            \\
            supergravity takes place here
            \end{tabular}
          }
        }
      }
    }{
      \mathrm{SupSmthGrpd}_\infty
    }
  \end{tikzcd}
  \hspace{5pt}
  }
  \\[-9pt]
  \\
  {\bf Figure 1.}
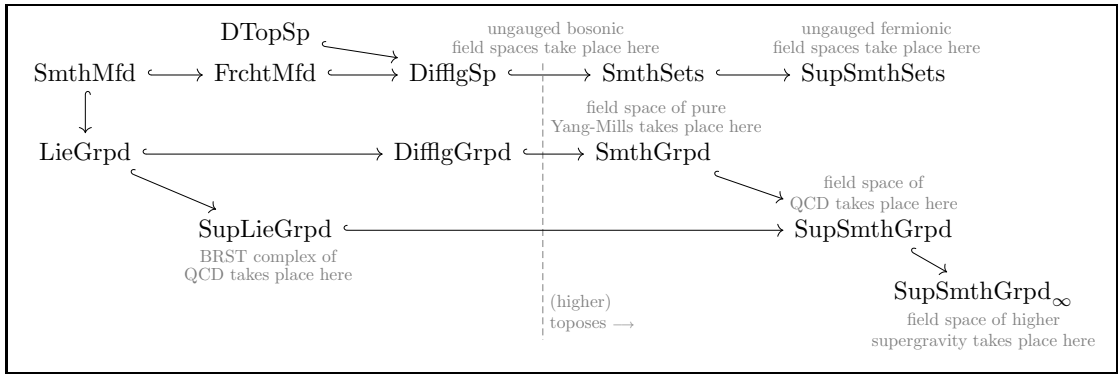
  Part of the system of categories of generalized spaces needed in mathematical physics.
  \end{tabular}
\end{strip}

\paragraph{Where variational calculus takes place.}
Informal physics texts often refer to variables whose value is non-vanishing but ``so tiny'' that their square may be neglected, $\epsilon^2 = 0$. This naive coordinate expressions -- which algebraically means that $\epsilon \in \mathbb{R}[\epsilon]/(\epsilon^2)$ -- may be understood as describing simple infinitesimal {\it probe} spaces, from which topos theory immediately provides us with rigorous models of global differential geometry containing actual infinitesimal quantities (historically known as ``synthetic differential geometry''):

Namely, a fundamental (if maybe underappreciated) fact of differential geometry is that smooth functions between smooth manifolds are {\it fully faithfully} reflected in the homomorphisms of real algebras of smooth functions which they induce:
\[
  \hspace{-2pt}
  \begin{tikzcd}[
    sep=3pt,
    row sep=-3pt
  ]
    \mathrm{Hom}_{\mathrm{SmthMfd}}\big(
      X
      ,\,
      Y
    \big)
    \ar[
      rr,
      phantom,
      "{ \simeq }"
    ]
    &&
    \mathrm{Hom}_{\mathrm{CAlg}_{\mathbb{R}}}\big(
      C^\infty(Y)
      ,\,
      C^\infty(X)
    \big)
    \\
    f
    \ar[
      rr,
      shorten=25pt,
      <->
    ]
    &&
    f^\ast
    \mathrlap{\,}
  \end{tikzcd}
\]
exhibiting a full embedding into the opposite category of commutative real algebras:
\[
  \begin{tikzcd}[
    column sep=15pt,
    row sep=-2pt
  ]
    \mathrm{CrtSp}
    \ar[r, hook]
    &
    \mathrm{SmthMnfd}
    \ar[
      rr,
      hook,
      "{
        C^\infty(-)
      }"
    ]
    &&
    \mathrm{CAlg}_{{}_{\mathbb{R}}}^{\mathrm{op}}
    \\
    &
    X
    \ar[
      rr,
      phantom,
      "{ \mapsto }"
    ]
    &&
    C^\infty(X)
    \mathrlap{\,.}
  \end{tikzcd}
\]
Reading this backwards -- just as familiar from algebraic geometry -- we may declare {\it infinitesimally thickened} Cartesian spaces to be that whose algebras of smooth functions, by definition, contain nilpotent monomials, defined as forming this full subcategory:
\begin{equation}
  \label{ThCartSp}
  \begin{tikzcd}[
    column sep=6pt,
    row sep=-2pt
  ]
    \hspace{-.3cm}
    \mathrm{ThCrtSp}
    \ar[
      rr,
      hook,
      "{
        C^\infty(-)
      }"
    ]
    &&
    \mathrm{CAlg}^{\mathrm{op}}_{\mathbb{R}}
    \\
    \mathbb{R}^n
    \!\times\!
    \mathbb{D}^m_k
    \ar[
      rr,
      phantom,
      "{ \mapsto }"
    ]
    &&
    C^\infty(\mathbb{R}^n)
    \otimes_{{}_{\mathbb{R}}}
    \mathbb{R}[
      \epsilon_1,
      \cdots,
      \epsilon_m
    ]\big/
    \big(
      \epsilon^{k+1}
    \big)
    \hspace{-.3cm}
  \end{tikzcd}
\end{equation}
which becomes a site via coverings of the form
\\
$
  \big\{
    U_i \times \mathbb{D}^{m}_k
    \xrightarrow{
      \iota_i \times \mathrm{id}
    }
    \mathbb{R}^n
     \times
    \mathbb{D}^{m}_k
  \Big\}_{i \in I}
$
for $(\iota_i)_{i \in I}$ a differentiably good open cover as before. Therefore we obtain \eqref{SheafTopos} the topos
\begin{equation}
  \label{SyntheticDifferentialTopos}
  \mathrm{FrmlSmthSet}
  \;:=\;
  \mathrm{Sh}\big(
    \mathrm{ThCrtSp}
    ,
    \mathrm{Set}
  \big)
\end{equation}
of (synthetic-){\it differential smooth sets}.
Smooth manifolds $X$ are still faithfully included here, by defining their plots algebraically
\begin{equation}
  \label{ToposOfInfinitesimallyThickenedSmthSets}
  \def\arraystretch{1.4}
  \begin{array}{l}
  \mathrm{Plt}\big(
    \mathbb{R}^n
    \times
    \mathbb{D}^m_k
    ,
    X
  \big)
  \\
  \,:=\,
  \mathrm{Hom}_{\mathrm{CAlg}_{\mathbb{R}}}\Big(
    C^\infty(X)
    ,\,
    C^\infty(
      \mathbb{R}^n
      \times
      \mathbb{D}^m_k
    )
  \Big)
  \,,
  \end{array}
\end{equation}
and hence exist now alongside
the infinitesimal halos $\mathbb{D}^m_{k}$. Notice that these contain only a single actual point
\[
  \iota
  :
  \ast
  =
  \mathbb{R}^0
  \xrightarrow{ \exists ! }
  \mathbb{D}^m_k
\]
and yet are larger than that point (have more plots, namely by other infinitesimals).

This topos \eqref{ToposOfInfinitesimallyThickenedSmthSets} is a most convenient environment where naive differential geometric intuition becomes a rigorous reality.
Eg., for $X$ a smooth manifold, its tangent bundle $T X \xrightarrow{p} X$ (also a smooth manifold), is just the mapping space out of the infinitesimal interval:
\begin{equation}
  \begin{tikzcd}[
    column sep=10pt
  ]
    \mathrm{Map}\big(
      \mathbb{D}^1_1
      ,
      X
    \big)
    \ar[r, phantom, "{ \simeq }"]
    \ar[
      d,
      "{
        \mathrm{Map}(
        \iota, X
        )
      }"
    ]
    &
    T X
    \ar[d, "{ p }"]
    \\
    \mathrm{Map}(\ast, X)
    \ar[r, phantom, "{ \simeq }"]
    &
    X
    \mathrlap{\,.}
  \end{tikzcd}
\end{equation}
Proceeding in this spirit, one finds that traditional variational calculus on jet bundles, hence all of classical Lagrangian field theory, takes place in $\mathrm{FrmlSmthSet}$ where the notorious technical subtleties are naturally being taken care of.

\paragraph{Where classical fermion fields take place.}

\noindent
It is commonplace that operator products of fermionic {\it quantum} fields satisfy Clifford algebra relations,
but it may be less widely appreciated that (therefore) already classical fermion fields are ``anti-commuting variables'' subject to the algebraic relation
\begin{equation}
  \label{AnticommutationRelation}
  \psi
  \,
  \psi'
  \;=\;
  -
  \psi'
  \,
  \psi
  \,,
\end{equation}
which is {\it necessary} for the usual Dirac-Lagrangian of fermion fields to make algebraic sense
\[
  L_\psi(x)
  \;\propto\;
  \overline{\psi}(x) D \psi(x)
  \,
  \mathrm{dvol}
  \,,
\]
since otherwise it would be a total derivative. But beyond such formal algebraic manipulations, what {\it is} a classical fermion field?
Remarkably, the naive algebraic coordinate-manipulations common in physics are again perfectly valid as statements about fermionic {\it probe} spaces, from where topos theory takes automatic care of providing a good notion of general fermionic (field) spaces:


Concretely, a simple {\it finite-dimensional Cartesian} fermionic space with $n$ bosonic and $q$ fermionic coordinate functions -- a ``super-space'' $\mathbb{R}^{n\vert q}$ -- ought to be fully characterized by the fact that its ``algebra of smooth functions'' $C^\infty(\mathbb{R}^{n\vert q})$ is (defined to be) the $\mathbb{Z}/2$-graded commutative (super-commutative) algebra (we write $\mathrm{sCAlg}_{\mathbb{R}}$ for their category) which is the plain tensor product of the ordinary smooth functions on $\mathbb{R}^n$ with the ``odd functions'' on $\mathbb{R}^q$, the latter defined to form the Grassmann algebra $\wedge^{\!\bullet}(\mathbb{R}^q)^\ast$ of the linear dual space:
\begin{equation}
  \label{FunctionAlgebraOnCartesianSuperSpace}
  C^\infty(\mathbb{R}^{n\vert q})
  :=
  C^\infty(\mathbb{R}^n)
  \!\otimes_{{}_{\mathbb{R}}}\!
  \wedge^{\!\bullet} (\mathbb{R}^q)^\ast
  \in
  \mathrm{sCAlg}_{\mathbb{R}}
  \,.
\end{equation}
As before for bosonic infinitesimals \eqref{ThCartSp}, we may {\it define} smooth maps of Cartesian super-spaces to {\it be} the reverse homomorphisms of their super-commutative function algebras:
\[
  f:
    \mathbb{R}^{n \vert q}
    \xrightarrow{\;}
    \mathbb{R}^{n' \vert q'}
  \;\leftrightarrow\;
  f^\ast:
    C^\infty(\mathbb{R}^{n' \vert q'})
    \xrightarrow{\;}
    C^\infty(\mathbb{R}^{n \vert q})
    \,.
\]
This means to define the category of smooth Cartesian super-spaces as the full sub-category of the opposite of the category $\mathrm{sCAlg}_{{}_{\mathbb{R}}}$  of super-commutative algebras on those of the form \eqref{FunctionAlgebraOnCartesianSuperSpace}:
\begin{equation}
  \label{SuperCartesianSpaces}
  \begin{tikzcd}[sep=0pt]
    \mathrm{SupCrtSp}
    \ar[
      rr,
      hook,
      "{ C^\infty(-) }"
    ]
    &&
    \mathrm{sCAlg}_{\mathbb{R}}^{\mathrm{op}}
    \\
    \mathbb{R}^{n \vert q}
    \,=\,
    \mathbb{R}^n
    \!\times\!
    \mathbb{R}^{0\vert q}
    &\mapsto&
    C^\infty(\mathbb{R}^n)
    \!\otimes_{{}_{\mathbb{R}}}\!
    \wedge^{\!\bullet} (\mathbb{R}^q)^\ast
    \mathrlap{\,.}
  \end{tikzcd}
\end{equation}
Declaring the coverings of Cartesian super-spaces to be of the form
$
  \big\{
    U_i \times \mathbb{R}^{0\vert q}
    \xrightarrow{
      \iota_i \times \mathrm{id}
    }
    \mathbb{R}^n
     \times
    \mathbb{R}^{0\vert q}
  \Big\}_{i \in I}
$
for $(\iota_i)_{i \in I}$ a differentiably good open cover as before, we readily obtain the topos of super smooth sets:
\begin{equation}
  \mathrm{SupSmthSet}
  \;:=\;
  \mathrm{Sh}(
    \mathrm{SupCrtSp}
    ,\,
    \mathrm{Set}
  )
  \,.
\end{equation}

For example, given a smooth spinor bundle $\mathcal{S} \to X$, it becomes a super smooth set $\Pi\mathcal{S}$ by:
\begin{equation}
  \label{PlotsOfSuperSmootuFunctionSet}
  \mathrm{Plt}\big(
    \mathbb{R}^{n\vert q}
    ,\,
    \Pi\mathcal{S}
  \big)
  :=
  \big\{
    \phi
    :
    \mathbb{R}^n
    \to
    X
  \big\}
  \times
  \wedge^{\!q}
  \Gamma\big(
    \phi^\ast \mathcal{S}
  \big)
\end{equation}
from which one finds that spinor fields are the odd points in the super smooth function set out of $\Pi \mathcal{S}$:
\[
 \def\arraycolsep{7pt}
 \def\arraystretch{1.2}
 \begin{array}{rcl}
  \Phi_{\mathrm{even}}
  :
  \mathbb{R}^{0 \vert 0}
  \to
  \mathrm{Map}(
    \Pi\mathcal{S},
    \mathbb{R}
  )
  &\leftrightarrow&
  \phi
  \in
   C^\infty(X)
   \\
  \Phi_{\mathrm{odd}}
  :
   \mathbb{R}^{0 \vert 1}
   \to
   \mathrm{Map}(
     \Pi \mathcal{S},
     \mathbb{R}
   )
   &\leftrightarrow&
   \psi
   \in
     \Gamma(\mathcal{S}^\ast)
   \mathrlap{\,.}
  \end{array}
\]
But under the following faithful embedding of supercommutative algebras into commutative algebras of super sets, the super smooth function set \eqref{PlotsOfSuperSmootuFunctionSet} is identified with the Grassmann algebra on spinor fields:
\[
  \def\arraystretch{1.3}
  \def\arraycolsep{2pt}
  \begin{array}{rcl}
    \mathrm{sCAlg}(\mathrm{Set})_{\mathbb{R}}
    &\xhookrightarrow{\phantom{--}}&
    \mathrm{CAlg}(\mathrm{SupSmthSet})_{\underline{\mathbb{R}}}
    \\
    A
    &\mapsto&
    \underline{A}
    \,:\,
    \mathbb{R}^{n\vert q}
    \mapsto
    \big(
      A
        \otimes
     C^\infty(\mathbb{R}^{n\vert q})
    \big)_{\mathrm{even}}
    \\[-6pt]
    \mathclap{
      \hspace{-80pt}
      \scalebox{1}{
        in that
      }
    }
    \\
    \Gamma\big(
      \!
      \wedge^{\!\bullet}
      \mathcal{S}^\ast
    \big)
    &\mapsto&
    \underline{
    \Gamma\big(
      \!
      \wedge^{\!\bullet}
      \mathcal{S}^\ast
    \big)
    }
    \,\simeq\,
    \mathrm{Map}\big(
      \Pi\mathcal{S}
      ,\,
      \mathbb{R}
    \big)
  \end{array}
\]
This exhibits the anticommutation relation \eqref{FunctionAlgebraOnCartesianSuperSpace}.

\smallskip

\noindent
\paragraph{Where discrete gauge fields take place.}
We may read the {\it gauge principle} in physics as saying that in gauge field spaces the very notion of equality is relaxed: Two (plots/families \eqref{PlotsOfMappingSpace} of) gauge fields
\[
  \Phi,
  \Phi'
  \,:\,
  \mathbb{R}^{n\vert q}
  \xrightarrow{\phantom{---}}
  \mathrm{Fields}
\]
may be nominally distinct and yet identified via gauge transformations:
\begin{equation}
  \label{GaugePfGaugeTransformationOfFields}
  \begin{tikzcd}
    \mathbb{R}^{n \vert q}
    \ar[
      rr,
      bend left=40,
      "{ \Phi }"{description, name=s}
    ]
    \ar[
      rr,
      bend right=40,
      "{ \Phi' }"{description, name=t}
    ]
    \ar[
      from=s,
      to=t,
      shorten=2pt,
      Rightarrow,
      "{ g }",
      "{ \sim }"{sloped, yshift=+.5pt},
    ]
    &&
    \mathrm{Fields}
  \end{tikzcd}
\end{equation}
that may be composed
\begin{equation}
  \label{CompositionOfGaugeTransformations}
  \begin{tikzcd}
    \mathbb{R}^{n \vert q}
    \ar[
      rr,
      bend left=60,
      "{ \Phi }"{description, name=s}
    ]
    \ar[
      rr,
      bend right=60,
      "{ \Phi'' }"{description, name=t}
    ]
    \ar[
      from=s,
      to=t,
      bend left=20,
      Rightarrow,
      "{ h \circ g }"{sloped, swap, pos=.76},
      "{ \sim }"{sloped, pos=.2},
    ]
    \ar[
      rr,
      crossing over,
      "{
        \Phi'
      }"{description, name=i, pos=.27}
    ]
    \ar[
      from=s,
      to=i,
      shorten=2pt,
      Rightarrow,
      "{ g }"{swap, pos=.6},
      "{ \sim }"{sloped, pos=.4, yshift=-.5pt, swap}
    ]
    \ar[
      from=i,
      to=t,
      shorten=2pt,
      Rightarrow,
      "{ h }"{swap, pos=.2},
      "{ \sim }"{sloped, swap, pos=.35, yshift=-.5pt, swap},
    ]
    &&
    \mathrm{Fields}
  \end{tikzcd}
\end{equation}
and reversed
\[
  g : \Phi \Rightarrow \Phi'
  \;\;\;\;
  \mbox{implies}
  \;\;\;\;
  \left\{
  \def\arraystretch{1.2}
  \begin{array}{l}
  g^{-1} : \Phi' \Rightarrow \Phi
  ,
  \\
  g^{-1} \circ g
  \,=\,
  \mathrm{id}_{\Phi}
  \\
  g \circ g^{-1}
  \,=\,
  \mathrm{id}_{\Phi'}
  \mathrlap{\,.}
  \end{array}
  \right.
\]
This means that the plots of gauge field spaces no longer form plain sets, but {\it groupoids}: sets of objects with invertible maps (gauge transformations) making them a category \eqref{CompositionOfMorphisms}.

Furthermore, for {\it higher} gauge fields, two such gauge transformations, in turn, may be nominally distinct and yet identified by ``gauge-of-gauge transformations''
\begin{equation}
  \begin{tikzcd}
    \mathbb{R}^{n \vert q}
    \ar[
      rr,
      bend left=40,
      "{ \Phi }"{description, name=s}
    ]
    \ar[
      rr,
      bend right=40,
      "{ \Phi' }"{description, name=t}
    ]
    \ar[
      from=s,
      to=t,
      shorten=2pt,
      bend left=50,
      Rightarrow,
      "{ \sim }"{sloped, swap, yshift=-.5pt},
      "{\ }"{name=tt, swap}
    ]
    \ar[
      from=s,
      to=t,
      shorten=2pt,
      bend right=50,
      Rightarrow,
      "{ \sim }"{sloped, yshift=.5pt},
      "{\ }"{name=ss}
    ]
    \ar[
      from=ss,
      to=tt,
      phantom,
      "{
        \Rightarrow
      }"{scale=1.5},
      "{ \sim }"{yshift=4.5pt, scale=.7, pos=.2}
    ]
    \ar[
      rr,
      -,
      shorten <=20pt,
      shorten >=16pt,
      line width=.5pt,
      shift left=1.3pt
    ]
    &&
    \mathrm{Fields}
  \end{tikzcd}
\end{equation}
satisfying a 2-dimensional analog of composition and associativity, as schematically indicated here:
\begin{equation}
  \label{ThreeSimplex}
  \begin{tikzcd}[
    row sep=25pt,
    column sep=13
  ]
    \mathllap{
      \scalebox{.7}{
        \color{gray}
        $(\Delta^0)$
      }
    }
    \Phi
    &&
    \phantom{\Phi'}
    \\[-25pt]
    \mathllap{
      \scalebox{.7}{
        \color{gray}
        $(\Delta^1)$
      }
    }
    \Phi
    \ar[
      rr,
      "{ g }"
    ]
    &&
    \Phi'
    \\[-25pt]
    &
    |[alias=two]|
    \Phi'
    \ar[
      dr,
      "{ h }"
    ]
    \\
    \Phi
    \ar[
      ur,
      "{
        \mathllap{
          \scalebox{.9}{
            \color{gray}
            $(\Delta^2)$
            \;\;\;\,
          }
        }
        g
      }"
    ]
    \ar[
      rr,
      "{ h \circ g }"{swap},
      "{\ }"{name=onethree}
    ]
    &&
    \Phi''
    \ar[
      from=two,
      to=onethree,
      Rightarrow,
      shorten <=5pt,
      "{
        \mu(g,h)
      }"{description, pos=.6}
    ]
  \end{tikzcd}
  \begin{tikzcd}
    &[-10pt]&
    &[-50pt]
    |[alias=two]|
    \Phi'
    \ar[
      ddr,
      "{\ }"{name=twofour, swap}
    ]
    &[-5pt]
    \\
    \\
    |[alias=one]|
    \Phi
    \ar[
      rrrr,
      "{\ }"{name=onefour}
    ]
    \ar[
      drr,
      "{\  }"{name=onethree}
    ]
    \ar[
      uurrr,
      "{
        \mathllap{
          \scalebox{.9}{
            \color{gray}
            $(\Delta^3)$
          }
        }
      }"{yshift=15pt}
    ]
    &&&&
    |[alias=four]|
    \Phi'''
    \mathrlap{\,,}
    \\[-3pt]
    &&
    |[alias=three]|
    \Phi''
    \ar[
      urr,
      "{\ }"{name=one}
    ]
    \ar[
      from=two,
      to=onefour,
      Rightarrow,
      crossing over,
      shorten=5pt
    ]
    \ar[
      from=two,
      to=onethree,
      Rightarrow,
      shorten <=10pt,
      shorten >=1pt,
      crossing over
    ]
    \ar[
      from=three,
      to=onefour,
      Rightarrow,
      crossing over,
      shorten=5pt
    ]
    \ar[
      from=two,
      to=three,
      crossing over,
      "{\ }"{name=twothree}
    ]
    \ar[
      from=three,
      to=twofour,
      shorten <= 10pt,
      shorten >= 2pt,
      Rightarrow,
      crossing over,
    ]
  \end{tikzcd}
  \hspace{-3pt}
\end{equation}
and so on to ever higher order gauge transformations, now making the plots form {\it higher groupoids}.

Traditional physics literature describes this phenomenon mostly infinitesimally, where it is captured by the homological algebra of {\it BRST complexes}: Here the infinitesimal gauge transformations appear as ``ghost fields'' and the infinitesimal higher-gauge transformations as ``ghost-of-ghost fields''.

The topos theory for going beyond the infinitesimal higher gauge transformations is again via probes: We detect the nature of a higher gauge groupoid $\mathcal{X}$ by recording the system of sets
$\mathrm{Plt}(\Delta^n, \mathcal{X})$
of $n$-dimensional higher gauge transformations of the shape indicated for low $n$ in \eqref{ThreeSimplex} -- called the $n$-{\it simplices} $\Delta^n$. Equipped with the fairly evident cell-preserving maps, these $n$-simplices form a site (with trivial coverage) denoted $\Delta$ (the ``simplex category''), so that higher groupoids should be found in the corresponding (pre-)sheaf topos of ``simplicial sets'',
\[
  \mathrm{Sh}(\Delta, \mathrm{Set})_{\mathrm{Kan}}
  \hookrightarrow
  \mathrm{Sh}(
    \Delta
    ,
    \mathrm{Set}
  )
  \,,
\]
as those simplicial sets all whose higher gauge transformations have composites, if suitably consecutive, in a manner that is suitably associative, unital and invertible up to further gauge transformations. All this turns out to be neatly encoded by the {\it Kan condition} which demands simply that whenever we find (probe) in a simplicial set $\mathcal{X}$ the boundary of an $n$-simplex {\it except} the $k$th $(n-1)$-face  -- called a {\it horn} $\Lambda^n_k \hookrightarrow \Delta^n$ -- then we may also find the missing face and the interior $n$-morphism:
\vspace{-.3cm}
\[
  \mathcal{X}
  \,\in\,
  \mathrm{Sh}(
    \Delta
    ,\,
    \mathrm{Set}
  )_{\mathrm{Kan}}
  \;\;\;\;\;
  \Leftrightarrow
  \;\;\;\;\;
  \begin{tikzcd}[
    row sep=9pt
  ]
    \mathllap{ \forall\;\; }
    \Lambda^n_k
    \ar[r]
    \ar[d, hook]
    &
    \mathcal{X}
    \\
    \Delta^n
    \mathrlap{\,.}
    \ar[
      ur,
      dashed,
      "{ \exists }", swap
    ]
  \end{tikzcd}
\]
Some important examples:

\noindent
(1.) Given a {\it discrete} group $G$, there is a groupoid
\begin{equation}
  \label{DeloopingGroupoid}
  \mathbf{B}G
  \;\;
  =
  \;\;
  \adjustbox{raise=10pt}{{``\adjustbox{raise=-10pt}{$\left\{
  \adjustbox{raise=-6pt}{
  \begin{tikzcd}
    \mathrm{pt}
      \ar[out=180-55, in=55,   looseness=4.5,
      "\scalebox{1}{$
        \;
        \mathclap{
          g
        }
        \;
      $}"{description},shift right=1]
  \end{tikzcd}
  }
  \,\middle\vert\;
  g \in G
  \right\}$}''}}
\end{equation}
with a single $\Delta^0$-plot $\mathrm{pt}$, one $\Delta^1$-plot $\mathrm{pt} \xrightarrow{g} \mathrm{pt}$ for each group element, $\Delta^2$-plots given by pairs of group elements and witnessing the group operation
\begin{equation}
  \label{PlotsOfDeloopingGroupoid}
  \mathrm{Plt}(
    \Delta^2,
    \,
    \mathbf{B}G
  )
  =
  \left\{
  \hspace{-5pt}
  \adjustbox{raise=4pt}{
  \begin{tikzcd}[
    column sep=5pt
  ]
    &
    |[alias=two]|
    \mathrm{pt}
    \ar[
      dr,
      "{ g_2 }"
    ]
    \\
    \mathrm{pt}
    \ar[rr]
    \ar[
      ur,
      "{ g_1 }"
    ]
    \ar[
      rr,
      "{
        g_2 \cdot g_1
      }"{swap},
      "{  }"{name=onethree}
    ]
    &&
    \mathrm{pt}
    \ar[
      from=two,
      to=onethree,
      Rightarrow,
      shorten <=5pt,
      "{ \exists ! }"
    ]
  \end{tikzcd}
  }
  \hspace{-5pt}
  \middle\vert
  \hspace{-3pt}
  \def\arraystretch{1.2}
  \begin{array}{c}
    (g_1, g_2)
    \\
    \in  G \!\times\! G
  \end{array}
  \hspace{-5pt}
  \right\}
  \,,
\end{equation}
and generally $
  \mathrm{Plt}(
    \Delta^n
    ,
    \mathbf{B}G
  )
  \,=\,
  G^{\times^n}
$.
This groupoid appears as the field fiber $F$ of $G$-Dijkgraaf-Witten theory.

\noindent
(2.) For a topological space $\mathrm{X} \in \mathrm{Top}$, its singular simplicial complex is Kan, representing the {\it higher path groupoid} $\shape \, \mathrm{X}$ (also ``fundamental''- or ``Poincar{\'e}''-groupoid):
\begin{equation}
  \label{PathInfinityGroupoid}
  \def\arraystretch{1.4}
  \begin{array}{l}
  \hspace{-.2cm}
  \mathrm{Plt}
  \big(
    \Delta^n,
    \shape \, \mathrm{X}
  \big)
  \\
  :=
  \mathrm{Hom}_{\mathrm{Top}}
  \Big(
    \big\{
      \vec x \in
      (\mathbb{R}_{\geq 0})^{n+1}
      \,\big\vert\,
      \textstyle{\sum}_i x_i = 1
    \big\}
    ,\,
    \mathrm{X}
  \Big)
  \,.
  \hspace{-.2cm}
  \end{array}
\end{equation}

\noindent
(3.) Given a chain-complex in non-negative degrees
\begin{equation}
 \label{ChainComplexInNonNegativeDegrees}
  V_\bullet
  =
  \big[
  \begin{tikzcd}[
    column sep=14pt
  ]
    \cdots
    \ar[r, "{ \partial_2 }"]
    &
    V_2
    \ar[r, "{ \partial_1 }"]
    &
    V_1
    \ar[r, "{ \partial_0 }"]
    &
    V_0
  \end{tikzcd}
  \big]
\end{equation}
it becomes a Kan-simplicial set $H V$ by
\begin{equation}
  \label{DoldKanConstruction}
  \mathrm{Plt}\big(
    \Delta^n
    ,
    H V
  \big)
  :=
  \mathrm{Hom}_{\mathrm{Ch}_\bullet}
  \Big(
    N_\bullet \mathbb{Z}[\Delta^n]
    ,\,
    V_\bullet
  \Big)
  \,,
\end{equation}
where we probe $V_\bullet$ with the normalized chains complex $N_\bullet(-)$ of cellular singular chains on $\Delta^n$.

In particular, when $V_\bullet = A[n]$ is concentrated on an abelian group in degree $n$, then
\begin{equation}
  \label{EilenbergMacLaneSpace}
  \mathbf{B}^n A
  \,:=\,
  H A[n]
  \,=:\,
  H(A,n)
\end{equation}
is known as the $n$th {\it Eilenberg-MacLane space} of $A$.

One readily checks that the maps between Kan simplicial sets form themselves Kan simplicial sets
\[
  \mathcal{X}
  ,\,
  \mathcal{Y}
  \,\in\,
  \mathrm{Sh}(\Delta)_{\mathrm{Kan}}
  \;\;
  \Rightarrow
  \;\;
  \mathrm{Map}(
    \mathcal{X}
    ,\,
    \mathcal{Y}
  )
  \,\in\,
  \mathrm{Sh}(\Delta)_{\mathrm{Kan}}
  \,.
\]
For example, the maps from $\shape \, \mathrm{X}$ \eqref{PathInfinityGroupoid} to $\mathbf{B}G$ \eqref{DeloopingGroupoid} form the groupoid of gauge fields and their gauge transformations in Dijkgraaf-Witten theory on $X$:
\[
  \mathrm{Map}\big(
    \shape \, X
    ,\,
    \mathbf{B}G
  \big)
  \;\;
  \simeq
  \;\;
  \mathrm{Flat}G\mathrm{Bund}_X
  \,.
\]
Moreover, these higher mapping groupoids have canonical composition operations
\[
  (\mbox{-})
  \circ
  (\mbox{-})
  :
  \mathrm{Map}(
    \mathcal{X}
    ,\,
    \mathcal{Y}
  )
  \times
  \mathrm{Map}(
    \mathcal{Y}
    ,\,
    \mathcal{Z}
  )
  \to
  \mathrm{Map}(
    \mathcal{X}
    ,\,
    \mathcal{Z}
  )
\]
which are associative and unital.
Hence in gauge-theoretic enhancement of the base topos of
\begin{itemize}
\item  sets

with sets of maps between them,
\end{itemize}
we have now
\begin{itemize}
\item Kan-simplicial sets

with Kan-simplicial sets of maps between them
\end{itemize}
behaving like higher (gauge) groupoids with higher groupoids of maps between them and forming what is called a category {\it enriched} in simplicial sets, to be denoted:
\begin{equation}
\label{EnrichedCategoryOfHigherGroupoids}
  \mathbf{Grpd}_\infty
  \;:=\;
  \mathbf{Sh}(
    \Delta,
    \mathrm{Set}
  )_{\mathrm{Kan  }}
  \,\in\,
  \mathrm{Sh}(\Delta)
  \mbox{-}\mathrm{Cat}
  \,.
\end{equation}
This is (one incarnation of) the default {\it higher topos} of {\it higher groupoids} or {\it higher homotopy types}.

In particular,
a
{\it gauge equivalence} (math jargon: {\it homotopy equivalence})
of higher groupoids
are maps
$\phi  : \mathcal{X}
\xleftrightarrow{\mathrm{heq}} \mathcal{X}' : \phi'$ which are inverses {\it up to gauge transformations} in that there exists:
\begin{equation}
  \label{HomotopyEquialence}
  \hspace{-.5cm}
  \begin{array}{c}
  g
  \,:\,
  \Delta^1
  \to
  \mathrm{Map}(\mathcal{X}, \mathcal{X})
  \,,
  \;\;\;
  g'
  \,:\,
  \Delta^1
  \to
  \mathrm{Map}(\mathcal{X}', \mathcal{X}')
  \\
  \begin{tikzcd}
    &
    |[alias=GPrime]|
    \mathcal{X}'
    \ar[
      dr,
      "{ \phi' }"
    ]
    \\
    \mathcal{X}
    \ar[rr, equals, "{\ }"{name=edge}]
    \ar[
      ur,
      "{ \phi }"
    ]
    &&
    \mathcal{X}
    \ar[
      from=GPrime,
      to=edge,
      Rightarrow,
      shorten <=2pt,
      "{ g }"{swap},
      "{ \sim }"{sloped}
    ]
  \end{tikzcd}
  \hspace{-20pt}
  \adjustbox{raise=-10pt}{
  \begin{tikzcd}
    \mathcal{X}'
    \ar[
      rr,
      equals,
      "{\ }"{swap, name=edge}
    ]
    \ar[
      dr,
      "{ \phi' }"{swap}
    ]
    &&
    \mathcal{X}'
    \\
    &
    |[alias=G]|
    \mathcal{X}
    \ar[
      ur,
      "{ \phi }"{swap}
    ]
    \\
    \ar[
      from=edge,
      to=G,
      Rightarrow,
      shorten >=2pt,
      "{ g' }"{swap},
      "{ \sim }"{sloped}
    ]
  \end{tikzcd}
  }
  \end{array}
  \hspace{-.5cm}
\end{equation}
Eg. there are homotopy equivalences
exhibiting \eqref{EilenbergMacLaneSpace}
as a {\it based loop space} (of basepoint-preserving maps and higher homotopies inside the mapping space)
\begin{equation}
  \label{DeloopingEquivalence}
  \hspace{-6pt}
  \begin{tikzcd}[
    column sep=8pt
  ]
  \mathbf{B}^n A
  \ar[
    r,
    shorten=-2pt,
    "{
      \mathrm{heq}
    }"{yshift=1pt}
  ]
  &
  \Omega_{\mathrm{pt}}
  \mathbf{B}^{n+1} A
  :=
  \mathrm{Map}^{\mathrm{pt}/\!}
  \big(
    \,
    \shape \, S^1
    ,
    \mathbf{B}^{n+1} A
    \,
  \big)
  \mathrlap{\,.}
  \end{tikzcd}
\end{equation}

One may equivalently understand the Kan-simplicial enrichment in \eqref{EnrichedCategoryOfHigherGroupoids} as the universal way of turning classes $W$ of maps that ought to be such homotopy equivalences \eqref{HomotopyEquialence}  -- but cannot be in an ordinary category -- into actual homotopy equivalences in a simplicially-enriched category, a process known as {\it simplicial localization} $\mathbf{L}^{\!W}$:
\begin{equation}
  \mathbf{Grpd}_\infty
  \;\underset{
    \scalebox{.6}{DK}
  }{\simeq}\;
  \mathbf{L}^{\!\mathrm{heq} }
  \,
  \mathrm{Sh}(\Delta)_{{}_{\mathrm{Kan}}}
  \,.
\end{equation}

\paragraph{Where smooth gauge fields take place.}
Via the paradigm of probes, it is now immediate that higher {\it (super-)smooth groupoids} $\mathcal{X}$ (faithfully subsuming Lie groupoids and diffeological groupoids) are whatever when probed with $\mathbb{R}^{n \vert q}$ exhibit a Kan-simplicial set of plots \eqref{GaugePfGaugeTransformationOfFields}:
\begin{equation}
  \label{SimplicialPresheaves}
  \def\arraystretch{1.4}
  \begin{array}{l}
  \mathrm{PSh}\big(
    \mathrm{SupCrtSp}
    ,\,
    \mathrm{Sh}(\Delta)_{\mathrm{Kan}}
  \big
  )
  \;=\;
  \\
  \left\{
  \hspace{13pt}
  \adjustbox{raise=2pt}{
  \begin{tikzcd}[
    row sep=-1pt,
    column sep=-4pt
  ]
    \mathllap{
      \mathcal{X}
      :
      \;
    } \mathrm{SupCrtSp}^{\mathrm{op}}
    \ar[
      r,
      shorten=-2pt
    ]
    &
    \mathrm{Sh}(\Delta)_{\mathrm{Kan}}
    \\
    \mathbb{R}^{n \vert q}
    \ar[
      r,
      phantom,
      "{ \mapsto }"
    ]
    &
    \mathbf{Plt}(\mathbb{R}^{n \vert q}, \mathcal{X})
    \\
  \end{tikzcd}
  }
  \right\}
  \end{array}
\end{equation}
In gauge-theoretic enhancement of \eqref{SmthSetsAsLocalizationOfPresheaves},
gauge equivalences between smooth higher groupoids
should be {\it local homotopy equivalences}:
\[
  \hspace{-13pt}
  \begin{tikzcd}[
    column sep=6pt
  ]
    \mathcal{X}
    \ar[
      r,
      shorten=-3pt,
      "{ \mathrm{lheq} }"{yshift=1pt},
      "{
        \scalebox{.7}{
          \color{gray}
          \def\arraystretch{.85}
          \begin{tabular}{c}
            local homotopy
            \\
            equivalence
          \end{tabular}
        }
      }"{swap, yshift=-2pt}
    ]
    &
    \mathcal{Y}
  \end{tikzcd}
  \hspace{-16pt}
  \Leftrightarrow
  \!
  \underset{
    n, q
  }{\forall}
  \hspace{-3pt}
  \begin{tikzcd}[
    column sep=5pt
  ]
    \mathbf{PltGrm}\big(
      \mathbb{R}^{n\vert q}
      ,
      X
    \big)
    \ar[
      r,
      shorten=-3pt,
      "{
        \mathrm{heq}
      }"{yshift=1pt},
      "{
        \scalebox{.7}{
          \color{gray}
          \def\arraystretch{.85}
          \begin{tabular}{c}
            homotopy
            \\
            equivalence
          \end{tabular}
        }
      }"{swap, yshift=-4pt}
    ]
    &
    \mathbf{PltGrm}\big(
      \mathbb{R}^{n\vert q}
      ,
      Y
    \big)
  \end{tikzcd}
\]
and the {\it higher topos} of (super) {\it smooth $\infty$-groupoids} (aka $\infty$-stacks, here incarnated as a simplicially enriched category) is the simplicial localization of \eqref{SimplicialPresheaves}:
\[
  \def\arraystretch{1.4}
  \begin{array}{l}
  \mathbf{SupSmthGrpd}_\infty
  \\
  :=
  \mathbf{L}^{\!\mathrm{lheq}}
  \,
  \mathrm{PSh}\Big(
    \mathrm{SupCrtSp}
    ,\,
    \mathrm{Sh}(\Delta, \mathrm{Set})_{\mathrm{Kan}}
  \Big)
  \end{array}
\]
Important examples:

\noindent
(1.) For a (super-)Lie group $G$, the analog of \eqref{PlotsOfDeloopingGroupoid} is
\begin{equation}
  \label{PlotsOfSuperSmoothDeloopingGroupoid}
  \def\arraystretch{1.2}
  \begin{array}{l}
  \mathrm{Plt}
  \Big(
    \Delta^2
    ,\,
    \mathbf{Plt}\big(
      \mathbb{R}^{n\vert q}
      ,
      \mathbf{B}G
    \big)
  \Big)
  \\
  =
  \left\{
  \hspace{-5pt}
  \adjustbox{raise=4pt}{
  \begin{tikzcd}[
    column sep=5pt
  ]
    &
    |[alias=two]|
    \mathrm{pt}
    \ar[
      dr,
      "{ g_2 }"
    ]
    \\
    \mathrm{pt}
    \ar[rr]
    \ar[
      ur,
      "{ g_1 }"
    ]
    \ar[
      rr,
      "{
        g_2 \cdot g_1
      }"{swap},
      "{  }"{name=onethree}
    ]
    &&
    \mathrm{pt}
    \ar[
      from=two,
      to=onethree,
      Rightarrow,
      shorten <=5pt,
      "{ \exists ! }"
    ]
  \end{tikzcd}
  }
  \hspace{-5pt}
  \middle\vert
  \hspace{-3pt}
  \def\arraystretch{1.2}
  \begin{array}{l}
    (g_1, g_2)
    \\
    \in
    C^\infty\big(
      \mathbb{R}^{n \vert q}
      ,\,
      G
    \big)^{\times^2}
  \end{array}
  \hspace{-5pt}
  \right\}
  \end{array}
\end{equation}
which exhibits the {\it super smooth groupoid} that deloops the Lie group $G$.

In slight variation, for $\mathfrak{g}$ denoting the (super-)Lie algebra of $G$ we now also have the smooth groupoid $\mathbf{B}G_{\mathrm{conn}}$ whose plots are $\mathfrak{g}$-valued connection forms $A$ (vector potentials) with their usual gauge transformations:
\[
  \def\arraystretch{1.6}
  \begin{array}{l}
    \mathrm{Plt}\big(
      \Delta^2
      ,
      \mathbf{Plt}(
        \mathbb{R}^{n\vert q}
        ,
        \mathbf{B}G_{\mathrm{conn}}
      )
    \big)
    \\
    :=
    \left\{
    \adjustbox{raise=4pt}{
    \def\arraystretch{1.4}
    \begin{tikzcd}[
      column sep=5pt
    ]
      &
      |[alias=two]|
      A_1
      \ar[
        dr,
        "{ g_2 }"
      ]
      \\
      A_0
      \ar[
        rr,
        "{
          g_2
          \cdot
          g_1
        }"{swap},
        "{\ }"{name=onethree}
      ]
      \ar[
        ur,
        "{
          g_1
        }"
      ]
      &&
      A_2
      \ar[
        from=two,
        to=onethree,
        shorten <=4pt,
        Rightarrow,
        "{ \exists! }"
      ]
    \end{tikzcd}
    }
    \,\middle\vert\,
    \def\arraystretch{1.2}
    \begin{array}{l}
      A_i \in
      \Omega^1(\mathbb{R}^{n \vert q})
      \otimes \mathfrak{g},
      \\
      g_i
      \in
      C^\infty(\mathbb{R}^{n \vert q}, G)
      \\
      A_i
      =
      g_i A_{i-1} g_i^{-1}
      \\
      \phantom{A_i =}
      +
      g_i
      \mathrm{d}
      g_i^{-1}
    \end{array}
    \right\}
  \end{array}
\]

\noindent
(2.) For $V_\bullet(-)$ a {\it sheaf} of chain complexes \eqref{ChainComplexInNonNegativeDegrees} on $\mathrm{SupCrtSp}$, such as the $d$th {\it Deligne complex}
\[
  \hspace{-1cm}
  \def\arraystretch{1.4}
  \begin{array}{l}
  \mathrm{Del}^d_\bullet
  \,:=\,
  \Big[
  \hspace{-4pt}
  \begin{tikzcd}[
    column sep=6pt
  ]
    \mathbb{Z}
    \ar[
      r,
      hook,
      shorten=-1pt
    ]
    &
    \Omega^0_{\mathrm{dR}}
    (-)
    \ar[
      r,
      shorten=-1.5pt,
      "{ \mathrm{d} }"{pos=.35}
    ]
    &
    \Omega^1_{\mathrm{dR}}
    (-)
    \ar[
      r,
      shorten=-1.5pt,
    ]
    &
    \cdots
    \ar[
      r,
      shorten=-1.5pt,
      "{
        \mathrm{d}
      }"{pos=.35}
    ]
    &
    \Omega^{d-1}_{\mathrm{dR}}(-)
  \end{tikzcd}
  \hspace{-4pt}
  \Big]
  \end{array}
  \hspace{-1cm}
\]
we get the (super-)smooth version $H V_\bullet(-)$ of the corresponding higher groupoid \eqref{DoldKanConstruction}, which we may denote
\[
  \mathbf{B}^d
  \mathrm{U}(1)_{\mathrm{conn}}
  \,:=\,
  H \mathrm{Del}^d
\]

\noindent
(3.) For $X$ a smooth manifold and any good open cover $\big\{ U_i \xhookrightarrow{\iota_i} X \big\}_{i \in I}$, its {\it {\v C}ech nerve} $\widehat X$ has as probes the smooth maps to the $U_i$ and as (higher) gauge transformations the maps into (higher) intersections:
\begin{equation}
  \label{PlotsOfCechNerve}
  \def\arraystretch{1}
  \begin{array}{l}
  \mathrm{Plt}\big(
    \Delta^k
    ,
    \mathbf{Plt}(
      \mathbb{R}^{n \vert q}
      ,
      \widehat X
    )
  \big)
  \\
  :=
  C^\infty\Big(
    \mathbb{R}^{n \vert q}
    ,
    \;
    \underset{
      \mathclap{
      i_1, \cdots, i_k
      }
    }{\coprod}
    \;
    U_{i_1}
    \cap
    \cdots
    \cap
    U_{i_k}
  \Big)
  \,.
  \end{array}
\end{equation}
This is locally homotopy equivalent to $X$:
\[
  \begin{tikzcd}[
    column sep=3pt,
    row sep=0pt
  ]
    \widehat{X}
    \ar[
      rr,
      "{ \mathrm{lheq} }"
    ]
    &&
    X
    \\
    \big(
    \mathbb{R}^{n\vert q}
    \to
    U_{i_1}
    \big)
    \ar[
      rr,
      phantom,
      "{ \mapsto }"
    ]
    &&
    \big(
    \mathbb{R}^{n\vert q}
    \to
    U_{i_1}
    \hookrightarrow X
    \big)
  \end{tikzcd}
\]
and hence a gauge-equivalent incarnation of $X$, but it is a ``good'' (namely {\it projectively cofibrant}) representative, implying that the mapping spaces out of $\widehat X$ into the above classifying objects $\mathbf{B}(-)_{(-)}$ exhaust the gauge equivalence classes of the corresponding maps.

The maps between these objects  are (modulate) (higher) gauge fields, classified by higher cohomology:
\begin{equation}
  \hspace{-2pt}
  \def\arraycolsep{.5pt}
  \begin{array}{|l|}
  \hline
  \begin{tikzcd}[
    column sep=9pt,
    row sep=3pt
  ]
  \mathrm{Map}\big(
    \widehat{X}
    ,
    \mathbf{B}G_{\mathrm{conn}}
  \big)
  \ar[d]
  \ar[
    r,
    phantom,
    "{ = }"
  ]
  &
  \bigg\{
  \hspace{-10pt}
  \mbox{
      \def\arraystretch{1}
      \begin{tabular}{c}
      $G$-Yang-Mills fields
      \\
      ($A$-fields)
      \end{tabular}
  }
  \hspace{-10pt}
  \bigg\}
  \ar[
    d,
    shorten <=-2pt,
    shorten >=-6pt
  ]
  \\
  \underbrace{
  \pi_0
  \mathrm{Map}\big(
    \widehat{X}
    ,
    \mathbf{B}G
  \big)
  }_{
    = H^1(X;\, G)
  }
  \ar[
    r,
    phantom,
    "{ = }"
  ]
  &
  \Big\{
  \hspace{-2pt}
  \mbox{
      $G$-instanton sectors
  }
  \hspace{-2pt}
  \Big\}
  \end{tikzcd}
  \\
  \hline
  \begin{tikzcd}[
    column sep=9pt,
    row sep=3pt
  ]
  \mathrm{Map}\big(
    \widehat{X}
    ,
    \mathbf{B}^2
    \mathrm{U}(1)_{\mathrm{conn}}
  \big)
  \ar[
    d
  ]
  \ar[
    r,
    phantom,
    "{ = }"
  ]
  &
  \Big\{
  \hspace{-2pt}
  \mbox{
    $B$-fields
  }
  \hspace{-2pt}
  \Big\}
  \ar[
    d,
    shorten <=-6pt
  ]
  \\
  \underbrace{
  \pi_0
  \mathrm{Map}\big(
    \widehat{X}
    ,
    \mathbf{B}^2
    \mathrm{U}(1)
  \big)
  }_{
    =\,  H^3(X;\, \mathbb{Z})
  }
  \ar[
    r,
    phantom,
    "{ = }"
  ]
  &
  \Big\{
  \hspace{-2pt}
  \mbox{
    \hspace{-10pt}
    \def\arraystretch{.9}
    \begin{tabular}{c}
      string
      \\
      charge sectors
    \end{tabular}
    \hspace{-10pt}
  }
  \hspace{-2pt}
  \Big\}
  \end{tikzcd}
  \\
  \hline
  \begin{tikzcd}[
    column sep=9pt,
    row sep=3pt
  ]
  \mathrm{Map}\big(
    \widehat{X}
    ,
    \mathbf{B}^3
    \mathrm{U}(1)_{\mathrm{conn}}
  \big)
  \ar[d]
  \ar[
    r,
    phantom,
    "{ = }"
  ]
  &
  \Big\{
  \hspace{-2pt}
  \mbox{
    $C$-fields
  }
  \hspace{-2pt}
  \Big\}
  \ar[
    d,
    shorten <=-6pt
  ]
  \\
  \underbrace{
  \pi_0
  \mathrm{Map}\big(
    \widehat{X}
    ,
    \mathbf{B}^2
    \mathrm{U}(1)
  \big)
  }_{
    =\,
    H^4(X;\, \mathbb{Z})
  }
  \ar[
    r,
    phantom,
    "{ = }"
  ]
  &
  \Big\{
  \hspace{-2pt}
  \mbox{
    \hspace{-10pt}
    \def\arraystretch{.9}
    \begin{tabular}{c}
      membrane
      \\
      charge sectors
    \end{tabular}
    \hspace{-10pt}
  }
  \hspace{-2pt}
  \Big\}
  \end{tikzcd}
  \\
  \hline
  \end{array}
\end{equation}

Beyond these examples, consider the countably infinite-dimensional complex Hilbert space $\HilbertSpace{H}$ with its topological space of Fredholm operators $\mathrm{Frd}(\HilbertSpace{H})$, which is a classifying space for topological K-theory. Then a cocycle in {\it differential} K-theory is a homotopy in $\mathbf{SmthGrpd}_\infty$ of the following form
\vspace{-3pt}
\begin{equation}
  \label{DifferentialKTheory}
  \begin{tikzcd}[
    row sep=2pt,
    column sep=20pt
  ]
    &
    \prod_k
    \boldsymbol{\Omega}
      ^{2k}
      _{\mathrm{clsd}}
    \ar[
      dd,
      Rightarrow,
      shorten=2pt,
      "{
        \scalebox{.7}{
          \begin{tabular}{c}
            RR-field
          \end{tabular}
        }
      }"{swap}
    ]
    \ar[
      dr,
      shorten=-4pt,
      "{
        \scalebox{.7}{
          \color{gray}
          \def\arraystretch{.8}
          \begin{tabular}{c}
            de Rham
            \\
            map
          \end{tabular}
        }
      }"{sloped}
    ]
    &[-15pt]
    \\
    \scalebox{.7}{
      \color{gray}
      \def\arraystretch{.9}
      \begin{tabular}{c}
        smth
        \\
        mfd
      \end{tabular}
    }
    \widehat{X}
    \ar[
      ur,
      dashed,
      shorten <= -3pt,
      shorten >=-3pt,
      "{
        \scalebox{.7}{
          \color{gray}
          \def\arraystretch{.9}
          \begin{tabular}{c}
            RR-flux
            \\
            densities
          \end{tabular}
        }
      }"{sloped, pos=.32}
    ]
    \ar[
      dr,
      dashed,
      shorten <= -5pt,
      shorten >= -2pt,
      "{
        \scalebox{.7}{
        \color{gray}
        \def\arraystretch{.9}
         \begin{tabular}{c}
           K-cocycle
         \end{tabular}
        }
      }"{sloped, swap, pos=.3}
    ]
    &
    &
    \prod_k
    \mathbf{B}^{2k} \, \mathbb{R}^\flat
    \\
    &
    \shape \, \mathrm{Frd}(\HilbertSpace{H})
    \ar[
      ur,
      shorten <=-3pt,
      shorten >=-2pt,
      "{
        \scalebox{.7}{
          \color{gray}
          \def\arraystretch{.9}
          \begin{tabular}{c}
            Chern
            \\
            scharacter
          \end{tabular}
        }
      }"{swap, sloped}
    ]
    \\
    {}
  \end{tikzcd}
\end{equation}
\vspace{-.5cm}

\noindent
Generally, every notion of generalized {\it differential cohomology} (classifying generalized higher gauge fields)
takes place in $\mathbf{SmthGrpd}_\infty$ in an analogous way.

\paragraph{Where singularities take place.} While smooth, spaces in physics famously may have singularities, by which we shall mean orbi-singularities:
A cone such as the quotient $\mathbb{R}^2/(\mathbb{Z}/n)$ of the plane (by rotation along an angle $2\pi/n$ about the origin) is smooth everywhere except at the tip, where the $\mathbb{Z}/n$-action appears to have ``shrunken away'' to act only ``inside the singular point'', appearing much like the groupoid $\mathbf{B}\mathbb{Z}/n$ \eqref{DeloopingGroupoid}.
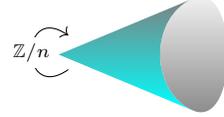
\[
  \begin{tikzcd}
    \phantom{\mathrm{pt}}
      \ar[
        out=180-55+90,
        in=55+90,
        looseness=4.5,
      "\scalebox{1}{$
        \;
        \mathclap{
          \mathbb{Z}/n
        }
        \;
      $}"{description},shift right=1]
  \end{tikzcd}
  \hspace{-23pt}
\adjustbox{
 raise=-.92cm
}{
\begin{tikzpicture}
\begin{scope}[yscale=.25, xscale=.6]
 \shadedraw[draw opacity=0, top color=gray, bottom color=cyan]
   (0,0) -- (3,3) .. controls (2,2) and (2,-2) ..  (3,-3) -- (0,0);
 \draw[draw opacity=0, top color=white, bottom color=gray]
   (3,3)
     .. controls (2,2) and (2,-2) ..  (3,-3)
     .. controls (4,-3.9) and (4,+3.9) ..  (3,3);
\end{scope}
\end{tikzpicture}
}
\]

Such an {\it orbifold} is hence a smooth set which moreover responds to probes by orbi-singularities, whose higher site may be understood to be the full simplicial subcategory on the delooping groupoids of finite groups:
\[
  \begin{tikzcd}[
    row sep=-4pt,
    column sep=10pt
  ]
    \mathbf{Snglr}
    \ar[
      rr,
      hook
    ]
    &&
    \mathbf{Grpd}_\infty
    \\
    \orbisingularG
    \ar[
      rr,
      phantom,
      "{ \mapsto }"
    ]
    &&
    \mathbf{B}G
  \end{tikzcd}
\]
With the gauge equivalences of orbi-singular (smooth) higher groupoids being the singularity-wise equivalences $\mathrm{sngeq}$ (notice that the trivial singularity is included as $\mathrm{pt} = \mathbf{B}1 \,\in\, \mathbf{Snglrt}$)
we obtain the higher topos of {\it singular} smooth higher groupoids:
\[
 \def\arraystretch{1.3}
 \begin{array}{l}
   \mathbf{SnglrSupSmthGrpd}_\infty
   \\
   :=
   \mathbf{L}^{\! \mathrm{sngeq} }
   \mathbf{PSh}\big(
     \mathbf{Snglrt}
     ,
     \mathbf{SupSmthGrpd}_\infty
   \big)
 \end{array}
\]
For $G \acts X$ a group action in smooth sets, the {\it homotopy quotient}
$X \!\sslash\! G$ is
\[
  \begin{array}{l}
  \mathrm{Plt}\Big(
    \Delta^2
    ,
    \mathbf{Plt}\big(
      \mathbb{R}^{n \vert q}
      ,
      X \!\sslash\! G
    \big)
  \Big)
  \\
  :=
  \left\{
  \adjustbox{raise=3pt}{
  \begin{tikzcd}[
    column sep=5pt
  ]
    &
    |[alias=two]|
    g_1 \cdot \phi
    \ar[
      dr,
      "{ g_2 }"
    ]
    &[-8pt]
    \\
    \phi
    \ar[
      ur,
      "{ g_1 }"
    ]
    \ar[
      rr,
      "{ g_2 \cdot g_1 }"{swap},
      "{\ }"{name=onethree}
    ]
    &&
    g_2 \!\cdot\! g_1 \cdot \phi
    \ar[
      from=two,
      to=onethree,
      Rightarrow,
      shorten <=4pt,
      end anchor={[xshift=3pt]},
      "{ \exists ! }"{pos=.48}
    ]
  \end{tikzcd}
  \hspace{-7pt}
  }
  \,\middle\vert\,
  \begin{array}{c}
    \phi \in
    \mathrm{Plt}\big(
      \mathbb{R}^{n \vert q}
      ,
      X
    \big)
    \\
    g_i \in
    \mathrm{Plt}\big(
      \mathbb{R}^{n \vert q}
      ,
      G
    \big)
  \end{array}
  \right\}
  \end{array}
\]
and its {\it orbi-singularization} $\orbisingular(-)$ is given by
\[
  \def\arraystretch{1.5}
  \begin{array}{l}
    \mathbf{Plt}\Big(
      \mathbb{R}^{n \vert q}
      \times
      \orbisingularG
      ,
      \orbisingular\big(
        X \!\sslash\! G
      \big)
    \Big)
    \\
    :=
    \mathbf{Plt}
    \Big(
    \mathbb{R}^{n \vert q}
    ,
    \mathrm{Map}\big(
      \mathbf{B}G
      ,\,
      X \!\sslash\! G
    \big)
    \Big)
  \end{array}
\]

For example, let $G \acts \HilbertSpace{H}$ be a stable representation of a finite group, then  a cocycle in the $G$-equivariant K-theory of $X$ is a map in $\mathbf{SnglrSmthGrpd}_\infty$ of this form:
\[
  \begin{tikzcd}[
    column sep=4pt,
    row sep=1pt
  ]
    \hspace{-.5cm}
    \orbisingular\big(
      X \!\sslash\! G
    \big)
    \ar[
      dr,
      shorten=-2pt,
      ->>
    ]
    \ar[
      rr,
      shorten=-2pt,
      "{
        \scalebox{.7}{
          \color{gray}
          \def\arraystretch{.9}
          \begin{tabular}{c}
            equivariant
            \\
            K-cocycle
          \end{tabular}
        }
      }"
    ]
    &&
    \shape
    \orbisingular\big(
      \mathrm{Frd}(\HilbertSpace{H})
        \!\sslash\!
      G
    \big)
    \ar[dl, ->>]
    \\
    &
    \orbisingularG
  \end{tikzcd}
  \hspace{-.5cm}
\]
From this one obtains, in analogy with \eqref{DifferentialKTheory}, equivariant {\it differential K-theory}, thought to modulate both RR-fields on orbifold spacetimes as well as ground states of topological phases of crystalline quantum materials (where the orbifold is the Brillouin torus quotiented by the point group of the crystal lattice).

\paragraph{Where quantum physics takes place.}
Indeed, physics is ultimately quantum, where quantum state spaces are {\it linear} spaces, varying over classical parameter spaces.

Linearity means {\it abelian group structure} and in higher gauge theory (homotopy theory), a higher group is the more abelian the more {\it de-loopings} \eqref{DeloopingEquivalence} it admits. Hence fully {\it linear structure} on a pointed higher groupoid $E_0$ should be a sequence of ever higher deloopings
(called a {\it spectrum})
exhibited by a sequence of maps
\[
  \begin{tikzcd}[
    column sep=20pt
  ]
    E_0
    \ar[
      r,
      "{ \tilde \sigma_0 }"
    ]
    &
    \Omega E_1
    \ar[
      r,
      "{
        \Omega(\tilde \sigma_1)
      }"{yshift=1pt}
    ]
    &
    \Omega^2 E_2
    \ar[
      r,
      "{
        \Omega^2(\tilde \sigma_2)
      }"{yshift=1pt}
    ]
    &
    \Omega^3 E_2
    \ar[
      r,
      "{
        \Omega^3(\tilde \sigma_3)
      }"{yshift=1pt}
    ]
    &
    \cdots
    \mathrlap{\,,}
  \end{tikzcd}
\]
which are homotopy equivalences up to suitable gauge equivalence of spectra, called {\it stable homotopy equivalences} ($\mathrm{steq}$).

Since maps into the looping of a higher groupoid are equivalently homotopies of this form,
\[
  X \xrightarrow{\;} \Omega_{\mathrm{pt}} Y
  \;\;\;\;\;
  \leftrightarrow
  \;\;\;\;\;
  \begin{tikzcd}[
    sep=6pt
  ]
    X
    \ar[r]
    \ar[d]
    &
    \mathrm{pt}
    \ar[d]
    \ar[
      dl,
      Rightarrow,
      shorten=-1pt,
      "{ \sim }"{sloped, pos=.35}
    ]
    \\
    \mathrm{pt}
    \ar[r]
    &
    Y
  \end{tikzcd}
\]
a parameterized spectrum is a space that may be probed by objects behaving like spheres of negative dimension, forming a higher site of this form:
\[
  \mathbf{Lin}
  :=
  \!
  \left\{
  \hspace{-9pt}
  \adjustbox{raise=2.5pt}{
  \begin{tikzcd}[
    column sep=5.5pt,
    row sep=6pt
  ]
  &
  \mathrm{pt}
  \ar[dl]
  \ar[from=dr]
  \ar[rr, equals]
  \ar[
    dd,
    shorten=5pt,
    Rightarrow,
    "{ \sim }"{sloped, pos=.44}
  ]
  &&
  \mathrm{pt}
  \ar[dl]
  \ar[from=dr]
  \ar[rr, equals]
  \ar[
    dd,
    shorten=5pt,
    Rightarrow,
    "{ \sim }"{sloped, pos=.44}
  ]
  &&
  \mathrm{pt}
  \ar[dl]
  \ar[from=dr]
  \ar[rr, equals, dashed]
  \ar[
    dd,
    shorten=5pt,
    Rightarrow,
    "{ \sim }"{sloped, pos=.44}
  ]
  &&
  {}
  \ar[dl, dashed]
  \\
  \mathbb{S}^0
  &&
  \mathbb{S}^{-1}
  &&
  \mathbb{S}^{-2}
  &&
  \mathbb{S}^{-3}
  &
  \\
  &
  \mathrm{pt}
  \ar[ul]
  \ar[from=ur]
  \ar[rr, equals]
  &&
  \mathrm{pt}
  \ar[ul]
  \ar[from=ur]
  \ar[rr, equals]
  &&
  \mathrm{pt}
  \ar[ul]
  \ar[from=ur]
  \ar[rr, equals, dashed]
  &&
  {}
  \ar[ul, dashed]
  \end{tikzcd}
  }
  \hspace{-9pt}
  \right\}
\]
Quite remarkably, parameterized spectra still form a higher topos, the higher {\it tangent topos}:
\begin{equation}
  \label{TangentTopos}
  \hspace{-1cm}
  \def\arraystretch{1.4}
  \begin{array}{l}
  \mathbf{LinSnglrSupSmthGrpd}_\infty
  \\
  :=
  \mathbf{L}^{\!\mathrm{steq} }
  \mathbf{PSh}\big(
    \mathbf{Lin}
    ,
    \mathbf{SnglrSupSmthGrpd}_{\infty}
  \big)
  \,.
  \end{array}
  \hspace{-1cm}
\end{equation}

In this higher topos \eqref{TangentTopos} takes place, for example:

\noindent
(1.)
The non-perturbative (``geometric'') quantization of  Poisson manifolds, exhibited here by the push-forward of K-module spectra parameterized (via a choice of higher {\it prequantum line bundle})
over the corresponding symplectic groupoid to its leaf space.

\noindent
(2.) The circuit logic of quantum information and quantum computing with classical control and dynamic lifting of quantum measurement results, exhibited by the base-change yoga of Real $H\mathbb{C}$-module spectra (as in Real K-theory), among whose ``heart'' are the finite-dimensional Hilbert spaces.

\noindent
(3.) The construction of Hilbert spaces of anyonic quantum ground states of  topologically ordered crystalline materials, in the guise of  $\mathfrak{su}(2)$-conformal blocks wih KZ-connection, exhibited here as sections of equivariant parameterized spectra over smooth configuration spaces of points in the crystal's Brillouin torus orbifold.

\medskip

\paragraph{Conclusion.} Quite contrary to superficial perception, higher topos theory provides just the mathematical context that physicists are often intuitively but informally assuming anyway. Realizing the higher topos theory yields a wealth of new powerful mathematical tools addressing many of the notoriously subtle issues in mathematical physics and opening the door to rigorous attacks on some of the outstanding open problems of the field.

There clearly remains a large gap between the languages that the communities speak, but this is a historical artefact which is increasingly being bridged.

However, this also means that, despite some history and considerable work, the application of higher topos theory in physics is still in its infancy. Our account here is by necessity a progress report more than the survey of mature field.

\vfill

{
  \footnotesize
  \hspace{-.6cm}
  \def\arraystretch{.9}
  \begin{tabular}{l}
  \rule{7.9cm}{.5pt}
  \\
  ${}^{\dagger}$
  Center for Quantum  and Topological Systems,
  \\
  \phantom{${}^{\dagger}$}
  Division of Science, New York University, Abu Dhabi,
  \\
  \phantom{${}^{\dagger}$} email: {\tt us13@nyu.edu}
  \\
  \rule{7.9cm}{.5pt}
  \end{tabular}
}

\noindent
\small
{\bf Keywords}:
 topos theory,
 higher topos theory, physics,
 mathematical physics,
  field theory, gauge theory, quantum theory, diffeology, groupoids, orbifolds, sheaves, stacks, higher structures

\newpage

\onecolumn

\noindent
{\large\bf References.}

\smallskip

\small

\noindent
{\bf Introductions to Category \& Topos theory.}

\reference
{Awodey S.}
{Category Theory}
{OUP}
{}
{}
{2006, 2010}
{}
{https://ncatlab.org/nlab/show/category+theory\#Awodey06}

\reference
{Geroch R.}
{Mathematical Physics}
{UCP}{}{}
{1985}
{}
{https://ncatlab.org/nlab/show/category+theory\#Geroch85}


\reference
{Schapira P.}
{An Introduction to Categories and Sheaves}
{lecture notes}
{}
{}
{2023}
{}
{https://ncatlab.org/nlab/show/Categories+and+Sheaves}

\reference
{Schreiber U.}
{Geometry of Physics: Categories and Toposes}
{nLab entry}
{}
{}
{2018-}
{}
{https://ncatlab.org/nlab/show/geometry+of+physics+--+categories+and+toposes}

\smallskip
\noindent
{\bf Diffeological Spaces and Smooth Sets.}

\reference
{Baez J. \& Hoffnung A.}
{Convenient Categories of Smooth Spaces}
{Trans. AMS}
{363}
{11}
{2011}
{}
{https://ncatlab.org/nlab/show/diffeological+space\#BaezHoffnung11}

\reference
{Iglesias-Zemmour P.}
{Diffeology}
{Mathematical Surveys \& Monographs, AMS}
{}
{}
{2013}
{}
{https://ncatlab.org/nlab/show/diffeological+space\#PIZ}

\reference
{Giotopoulos G. et al.}
{Smooth Sets of Fields}
{preprint}
{}
{}
{2023}
{}
{https://ncatlab.org/schreiber/show/Smooth+Sets+of+Fields}

\reference
{Schreiber, U.}
{Geometry of Physics: Smooth Sets}
{nLab entry}
{}
{}
{2014-}
{}
{https://ncatlab.org/nlab/show/geometry+of+physics+--+smooth+sets}

\smallskip

\noindent
{\bf Synthetic Differential Geometry over $\mathrm{ThCartSp}$.}

\reference
{Dubuc E.}
{Sur les mod{\`e}les de la g{\'e}om{\'e}trie diff{\'e}rentielle synth{\'e}tique}
{Cahiers}
{20}
{3}
{1979}
{231-279}
{http://www.numdam.org/item?id=CTGDC_1979__20_3_231_0}

\reference
{Kock A.}
{Convenient vector spaces embed into the Cahiers topos}
{Cahiers}
{27}
{1}
{1986}
{3-17}
{http://www.numdam.org/item?id=CTGDC_1986__27_1_3_0}

\reference
{Kock A. \& Reyes G.}
{Corrigendum and addenda to [Kock 1986]}
{Cahiers}
{28}
{2}
{1987}
{99-110}
{http://www.numdam.org/item?id=CTGDC_1987__28_2_99_0}

\reference
{Schreiber U. \& Khavkine I.}
{Synthetic Geometry of Differential Equations}
{preprint}
{}
{}
{2017}
{}
{https://arxiv.org/abs/1701.06238}

\smallskip

\noindent
{\bf Super Smooth Sets.}

\reference
{Konechny A. \& Schwarz A.}
{On $(k \oplus l/q)$-dimensional supermanifolds}
{LNP}
{509}
{}
{1998}
{}
{https://doi.org/10.1007/BFb0105247}

\reference
{Sachse C.}
{A Categorical Formulation of Superalgebra and Supergeometry}
{preprint}
{}
{}
{2008}
{}
{https://arxiv.org/abs/0802.4067}

\reference
{Schreiber U.}
{Geometry of Physics: Supergeometry}
{nLab entry}
{}
{}
{2016-}
{}
{https://ncatlab.org/nlab/show/geometry+of+physics+--+supergeometry}

\reference
{Yetter D.}
{Models for synthetic supergeometry}
{Cahiers}
{29}
{2}
{1988}
{}
{http://www.numdam.org/item?id=CTGDC_1988__29_2_87_0}

\smallskip

\noindent
{\bf Higher Category \& Topos Theory.}

\reference
{Jur{\v c}o B. et al}
{Higher Structures in M-Theory}
{Fort. Phys.}
{67}
{8-9}
{2019}
{}
{https://ncatlab.org/nlab/show/Higher+Structures+in+M-Theory+2018}

\reference
{Lurie, J.}
{Higher Topos Theory}
{Ann. Math Stud., PUP}
{170}
{}
{2009}
{}
{https://press.princeton.edu/titles/8957.html}

\reference
{Rezk S.}
{Lectures on Higher Topos Theory}
{lecture notes}
{}
{}
{2019}
{}
{https://ncatlab.org/nlab/files/RezkHigherToposTheory2019.pdf}

\reference
{Schreiber U.}
{Higher Prequantum Geometry}
{New Spaces for Mathematics, CUP}
{}
{}
{2021}
{}
{https://ncatlab.org/schreiber/show/Higher+Prequantum+Geometry}

\smallskip

\noindent
{\bf Smooth groupoids and Gauge theory.}

\reference
{Benini M. et al.}
{The stack of Yang-Mills fields on Lorentzian manifolds}
{CMP}
{359}
{}
{2018}
{765-820}
{https://link.springer.com/article/10.1007/s00220-018-3120-1}

\reference
{Pavlov D. \& Grady D.}
{The geometric cobordism hypothesis}
{preprint}
{}
{}
{2021}
{}
{https://arxiv.org/abs/2111.01095}

\reference
{Fiorenza D. et al.}
{A higher stacky perspective on CS-theory}
{Math. Aspects of QFT, Springer}
{}
{}
{2014}
{153-211}
{https://ncatlab.org/schreiber/show/A+higher+stacky+perspective+on+Chern-Simons+theory}

\smallskip

\noindent
{\bf Smooth $\infty$-Groupoids and Differential Cohomology.}

\reference
{Bunke U. et al.}
{Differential cohomology theories as sheaves of spectra}
{JHRS}
{11}
{}
{2016}
{1-66}
{https://ncatlab.org/nlab/show/differential+cohomology+diagram\#BunkeNikolausVoelkl13}

\reference
{Fiorenza D. et al.}
{{\v C}ech cocycles for differential characteristic classes}
{ATMP}
{16}
{}
{2012}
{149-250}
{https://ncatlab.org/schreiber/show/Cech+Cocycles+for+Differential+characteristic+Classes}

\reference
{Hopkins M. \& Quick G.}
{Hodge filtered complex bordism}
{J. Top.}
{8}
{1}
{2014}
{147-183}
{https://ncatlab.org/nlab/show/Hodge-filtered+differential+cohomology\#HopkinsQuick14}

\reference
{Hopkins M. \& Singer I.}
{Quadratic Functions in Geometry, Topology, \& M-Theory}
{JDG}
{70}
{3}
{2005}
{329-452}
{https://ncatlab.org/nlab/show/Quadratic+Functions+in+Geometry,+Topology,+and+M-Theory}

\reference
{Sati H. et al.}
{Twisted Differential String \& Fivebrane Structures}
{CMP}
{315}
{1}
{2012}
{169-213}
{https://ncatlab.org/schreiber/show/Twisted+Differential+String+and+Fivebrane+Structures}

\reference
{Sati H. et al.}
{The Character Map in Nonabelian Cohomology}
{World Scientific}
{}
{}
{2023}
{}
{https://doi.org/10.1142/13422}

\reference
{Schreiber U.}
{Differential Cohomology in a Cohesive $\infty$-Topos}
{preprint}
{}
{}
{2013}
{}
{https://ncatlab.org/schreiber/show/differential+cohomology+in+a+cohesive+topos}

\reference
{Schreiber U.}
{Geometry of Physics: Smooth Homotopy Types}
{nLab entry}
{}
{}
{2014-}
{}
{https://ncatlab.org/nlab/show/geometry+of+physics+--+categories+and+toposes}

\smallskip

\noindent
{\bf Singular smooth $\infty$-groupoids and Orbifold Cohomology.}

\reference
{Rezk C.}
{Global Homotopy Theory and Cohesion}
{note}
{}
{}
{2014}
{}
{https://ncatlab.org/nlab/show/Global+Homotopy+Theory+and+Cohesion}

\reference
{Sati H. \& Schreiber U.}
{Proper Orbifold Cohomology}
{preprint}
{}
{}
{2020}
{}
{https://ncatlab.org/schreiber/show/Proper+Orbifold+Cohomology}

\reference
{Sati H. \& Schreiber U.}
{Equivariant Principal $\infty$-Bundles}
{preprint}
{}
{}
{2021}
{}
{https://ncatlab.org/schreiber/show/Equivariant+principal+infinity-bundles}

\smallskip

\noindent
{\bf Linearized $\infty$-Groupoids and Quantum Physics.}

\reference
{Sati H. et al.}
{Topological Quantum Gates in Homotopy Type Theory}
{preprint}
{}
{}
{2023}
{}
{https://arxiv.org/abs/2303.02382}

\reference
{Sati H. \& Schreiber U.}
{Entanglement of Sections}
{preprint}
{}
{}
{2023}
{}
{https://ncatlab.org/schreiber/show/Entanglement+of+Sections}

\reference
{Sati H. \& Schreiber U.}
{The Quantum Monadology}
{preprint}
{}
{}
{2023}
{}
{https://ncatlab.org/schreiber/show/The+Quantum+Monadology}

\reference
{Schreiber U.}
{Quantization via Linear Homotopy Types}
{talk notes}
{}
{}
{2014}
{}
{https://ncatlab.org/schreiber/show/Quantization+via+Linear+homotopy+types}


\end{document}